\documentclass[reprint,aps,prl,amsmath,amssymb,superscriptaddress,longbibliography,nobibnotes,floatfix]{revtex4-2}
\usepackage{graphicx, amsmath, amssymb, color, xcolor, orcidlink, bm}
\usepackage{setspace}
\usepackage[utf8]{inputenc}
\usepackage{nicefrac}
\usepackage{multirow, makecell, ragged2e}
\usepackage[justification=raggedright, singlelinecheck=false]{caption}
\usepackage{adjustbox}
\usepackage{tikz}
\usetikzlibrary{arrows.meta, positioning}
\graphicspath{{./}}
\newcommand{\be}{\begin{equation}}
\newcommand{\ee}{\end{equation}}
\newcommand{\LLL}{\text{P}_{\text{LLL}}}
\newcommand{\conj}[1]{\bar{#1}}
\newcommand{\overlap}[2]{\langle #1 \vert #2 \rangle}
\newcommand{\Lrel}{L_{\mathrm{rel}}}
\newcommand{\prlhead}[1]{\textbf{#1}}

\newcommand{\lamf}[7]{%
  \begin{scope}[shift={(#1,0)}]
    \begin{scope}[local bounding box=partonlevels]
    \foreach \occ [count=\k from 0] in #4 {%
      \ifnum\k=#5\relax
        \ifnum#6>0\relax
          \pgfmathsetmacro{\bxa}{0.34*(\occ+1)-0.27}\pgfmathsetmacro{\bxb}{0.34*(\occ+#6)+0.07}%
        \else
          \pgfmathsetmacro{\bxa}{0.07}\pgfmathsetmacro{\bxb}{0.34*\occ+0.07}%
        \fi
        \fill[yellow!60,rounded corners=1.2pt]
          (\bxa,0.66*\k-0.27) rectangle (\bxb,0.66*\k+0.27);
        \draw[black!55,dashed,line width=0.6pt,rounded corners=1.2pt]
          (\bxa,0.66*\k-0.27) rectangle (\bxb,0.66*\k+0.27);
      \fi
      \draw[black!45,line width=2.3pt,line cap=round] (-0.04,0.66*\k) -- (1.80,0.66*\k);
      \ifnum\occ>0\relax
        \foreach \j in {1,...,\occ}{%
          \pgfmathsetmacro{\cx}{0.34*\j-0.10}%
          \draw[fill=#2,draw=black!75,line width=0.9pt] (\cx,0.66*\k) circle (0.158);
        }%
      \fi
      \ifnum\k=#5\relax\ifnum#6>0\relax
        \foreach \j in {1,...,#6}{%
          \pgfmathsetmacro{\cx}{0.34*(\occ+\j)-0.10}%
          \draw[draw=black!75,dashed,line width=0.9pt,fill=white] (\cx,0.66*\k) circle (0.158);
        }%
      \fi\fi
    }%
    \end{scope}
    \draw[#7,black,line width=1.35pt,>={Stealth[length=4.8pt,width=4.5pt]}]
      ([yshift=2.5pt]partonlevels.north) -- ++(0,0.57);
    \node[font=\small] at (0.88,-0.52) {#3};
  \end{scope}%
}

\ifdefined\pdfsuppresswarningpagegroup\pdfsuppresswarningpagegroup=1\fi

\begin{document}
\hypersetup{hidelinks}
\title{Search for free non-Abelian anyons in paired composite-fermion states}
\author{Mytraya Gattu~\orcidlink{0000-0001-6994-389X}}
\thanks{Current address: Research Laboratory of Electronics, Massachusetts Institute of Technology, Cambridge, Massachusetts 02139, USA; email: \texttt{mytraya@mit.edu}.}
\affiliation{Department of Physics, 104 Davey Lab, Pennsylvania State University, University Park, Pennsylvania 16802, USA}
\affiliation{Center for Theory of Emergent Quantum Matter, Pennsylvania State University, University Park, Pennsylvania 16802, USA}
\author{J. K. Jain~\orcidlink{0000-0003-0082-5881}}
\email{jkj2@psu.edu}
\affiliation{Department of Physics, 104 Davey Lab, Pennsylvania State University, University Park, Pennsylvania 16802, USA}
\affiliation{Center for Theory of Emergent Quantum Matter, Pennsylvania State University, University Park, Pennsylvania 16802, USA}
\affiliation{Lodha Theoretical Physics Institute, 17th Floor, Lodha NCP Supremus, Wadala (E), Mumbai 400037, India}

\date{\today}
\begin{abstract}
The even-denominator fractional quantum Hall (FQH) states, understood as 
topological superconductors of composite fermions, are predicted to support elementary quasiparticles (QPs) or quasiholes (QHs) obeying non-Abelian braid statistics. 
Using the microscopic framework of parton theory, we investigate many experimentally observed paired composite-fermion states and find that their elementary QPs and QHs in general bind to form Abelian molecules, thus obscuring non-Abelian braid statistics. An exception is the $\nu=1/2$ state in wide quantum wells, where our calculations show free QPs or QHs; we thus identify this state as an especially promising candidate for realizing and studying non-Abelian braid statistics. We also indicate how the study of such bound states can help reveal the topological structure of the state.
\end{abstract}
\maketitle

The first even-denominator fraction, $\nu=5/2$, was observed in 1987~\cite{Willett87}. It is believed to be a topological $p$-wave paired state of composite fermions (CFs), which may be modeled by a Pfaffian or an anti-Pfaffian wave function~\cite{Moore91, Read00,Levin07,Lee07}, a parton wave function~\cite{Balram18}, or a Bardeen-Cooper-Schrieffer (BCS) type wave function of CFs~\cite{Sharma21}, which all presumably constitute different approximations to the actual paired CF state. Furthermore, these have been predicted to host quasiparticles (QPs) and quasiholes (QHs) of charge $\pm e/4$ obeying non-Abelian braid statistics~\cite{Moore91,Read00,Stern04,Read09,Bonderson11}. Non-Abelian anyons have been proposed to be useful for fault-tolerant topological quantum computation~\cite{DasSarma05,DasSarma06,Stern06,Nayak08,Stern10,DasSarma15}. By now, a plethora of even-denominator states have been observed and modeled as paired CF states. These include: $\nu=1/2$ in wide GaAs quantum wells~\cite{Suen92, Shabani09, Sharma24}; $\nu=1/4$ in wide quantum wells (WQWs)~\cite{Shabani09, Shabani13, Sharma24}; $\nu=3/4,1/4$ in narrow quantum wells with large Landau-level (LL) mixing~\cite{Wang22A,Zhao23,Wang23}; $\nu=1/2$ in $n=2$ and $n=3$ LLs of monolayer graphene~\cite{Kim19,Sharma23,Domaretskiy26}; $\nu=1/2$ in bilayer graphene in Landau levels that are analogous to the first excited LL of GaAs quantum wells~\cite{Ki14, Zibrov17, Li17, Li18}; and there is evidence for fractional quantum Hall effect (FQHE) at $\nu=1/6$ and $\nu=1/8$~\cite{Wang25,Balram25}.  The general agreement between theory and experiment makes a compelling case for the CF-pairing as the underlying mechanism of these even-denominator states. In this context it is especially noteworthy that the theoretical phase diagram of the $1/2$ and $1/4$ fractional quantum Hall (FQH) states as a function of the density and width is in excellent agreement with the experimental phase diagram~\cite{Sharma24}; the $1/4$ FQHE in the presence of high LL mixing was predicted in Ref.~\cite{Zhao23} and observed in Ref.~\cite{Wang23}; and $1/2$ FQHE in the $n=2$ graphene LL was predicted in Ref.~\cite{Sharma23} and observed in Ref.~\cite{Domaretskiy26}. Table~\ref{tab:overview} summarizes the observed even-denominator states and their predicted CF pairing channels.

\begin{table*}[tp]
  \caption{Observed even-denominator states (first column), their parton representations (second column), and the composite-fermion (CF) pair angular momentum $\ell$ (third column). The fourth and fifth columns give the results for two quasiholes (QHs) and two quasiparticles (QPs) in the odd and even channels. $\Lrel^{\star}$ denotes the relative angular momentum of the lowest-energy bound state, and $\Delta_2$ is its binding energy in units of $e^2/(\epsilon\ell_{\rm B})$ for systems with $N\sim100$ electrons. LL denotes Landau level; MLG denotes monolayer graphene.
  The wide-quantum-well results are for $w=60$\,nm and $\rho=2.01\times10^{11}$\,cm$^{-2}$ (Fig.~\ref{fig:wlb}).
  Density profiles show the deviation of the local filling factor from its average, $\Delta\nu/\nu$ (see also Fig.~\ref{fig:density}). The last column lists the relevant theoretical references. At $\nu=1/2$, the results for anti-$\bar{2}^{2}1^{3}$ may be obtained by particle-hole transformation. For higher-LL states, the displayed profiles are lowest LL densities; the one-body form factor required to recover the density in the appropriate LL has not been applied. The chiral central charge for the paired state, $c=1+\ell/2$, can be measured in thermal Hall experiments.}
  \label{tab:overview}
  \label{fig:Table}
  \footnotesize
  \setlength{\tabcolsep}{2pt}
  \renewcommand{\arraystretch}{1.15}
  \newcommand{\overviewprofile}[1]{\includegraphics[width=0.11\textwidth]{#1}}
  \newcommand{\overviewprofilecrop}[1]{\includegraphics[page=1,viewport=#1,clip,width=0.11\textwidth]{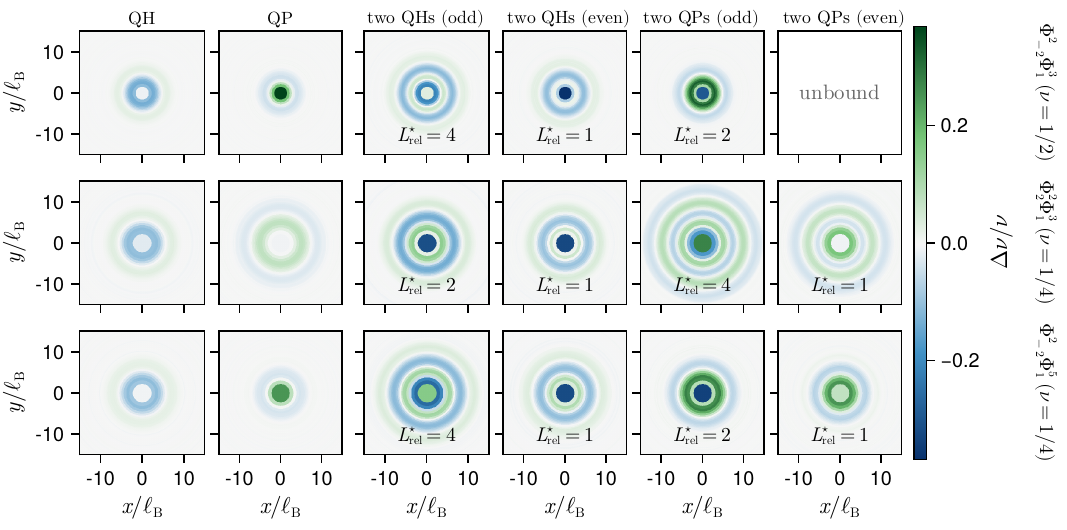}}
  \begin{ruledtabular}
  \begin{tabular}{l|c|c|c@{}c|c@{}c|c}
    State & Model & \makecell{Pair\\ang.mom.\\$\ell$} & \multicolumn{2}{c|}{Two QHs} & \multicolumn{2}{c|}{Two QPs} & Refs.\\
    \cline{4-7}
     & & & \makecell{Density\\profile} & (parity, $\Lrel^{\star}$, $\Delta_2 (e^2/\epsilon \ell_{\rm B})$) & \makecell{Density\\profile} & (parity, $\Lrel^{\star}$, $\Delta_2 (e^2/\epsilon \ell_{\rm B})$) & \\
    \hline
    \makecell[l]{$1/2$ in GaAs\\$n=1$ LL\\($\nu_{\rm total}=5/2$)\\\cite{Willett87}}
      & $\bar{2}^{2}1^{3}$ & $-3$ & \overviewprofile{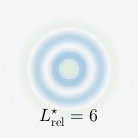} & $(\mathrm{odd}, 6,\sim\!{-}0.003)$ & \makecell{\overviewprofile{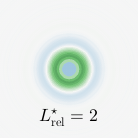}\\[-2pt]\overviewprofile{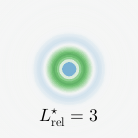}} & \makecell{$(\mathrm{odd}, 2,\sim\!{-}0.005)$\\$(\mathrm{even}, 3,\sim\!{-}0.005)$} & \makecell{\cite{Balram18}\\\cite{Sharma21}}\\
    \hline
    \makecell[l]{$1/2$ in wide\\GaAs wells\\$n=0$~\cite{Suen92,Shabani13}}
      & $\bar{2}^{2}1^{3}$ & $-3$ & -- & \makecell{free (within\\numerical uncertainty)} & \overviewprofile{density_profile_2bar2bar111_QP_Lrel2.pdf} & $(\mathrm{odd}, 2,\sim\!{-}0.006)$ & \makecell{\cite{Balram18}\\\cite{Sharma24}}\\
    \hline
    \makecell[l]{$1/4$ in wide\\GaAs wells\\$n=0$~\cite{Shabani09,Shabani09A}}
      & $2^{2}1^{3}$ & $+3$ & \makecell{\overviewprofilecrop{175 108 235 167}\\[-2pt]\overviewprofilecrop{241 108 301 167}} & \makecell{$(\mathrm{odd}, 2,\sim\!{-}0.001)$\\$(\mathrm{even}, 1,\sim\!{-}0.001)$} & \makecell{\overviewprofilecrop{308 108 367 167}\\[-2pt]\overviewprofilecrop{374 108 433 167}} & \makecell{$(\mathrm{odd}, 4,\sim\!{-}0.003)$\\$(\mathrm{even}, 1,\sim\!{-}0.002)$} & \makecell{\cite{Sharma24}\\\cite{Faugno19}}\\
    \hline
    \makecell[l]{$1/2$ in MLG\\$n=2$ LL~\cite{Domaretskiy26}}
      & $\bar{2}^{2}1^{3}$ & $-3$ & \overviewprofile{density_profile_2bar2bar111_QH_Lrel6.pdf} & $(\mathrm{odd}, 6,\sim\!{-}0.007)$ & \overviewprofile{density_profile_2bar2bar111_QP_Lrel3.pdf} & $(\mathrm{even}, 3,\sim\!{-}0.021)$ & \makecell{\cite{Balram18}\\\cite{Sharma23}}\\
    \hline
    \makecell[l]{$1/2$ in MLG\\$n=3$ LL~\cite{Kim19}}
      & $2^{2}1$ & $+3$ & -- & -- & -- & -- & \cite{Sharma23}\\
    \hline
    \makecell[l]{$1/4$ in GaAs\\holes\\strong LL\\mixing~\cite{Wang23}}
      & $\bar{2}^{2}1^{5}$ & $-3$ & -- & -- & -- & -- & \cite{Zhao23}\\
  \end{tabular}
  \end{ruledtabular}
  \par\smallskip
\end{table*}

The nature of the QP / QH excitations of these states remains unsettled at present.  At $\nu=5/2$, the predicted charge $e/4$ is seen at intermediate tunneling strengths~\cite{Dolev08}, while $e/2$ is observed
for weak tunneling and at low temperatures~\cite{Dolev10} in shot-noise experiments. In bilayer graphene, the interference period at even-denominator fillings corresponds to $e/2$~\cite{Kim26}, whereas shot noise at $\nu=1/2$ in a wide GaAs quantum well gives $e/4$~\cite{Alkalay26}. The charge $e/2$ has been attributed to a correlated tunneling of two QPs at once~\cite{Feldman21}, although the underlying mechanism is not fully understood. There has not been a universally accepted observation of non-Abelian braid statistics either. Theoretically, the systems accessible to exact diagonalization or other methods are too small to capture the physics of the QPs / QHs of the paired states quantitatively, as they are generally quite large and, moreover, always come in pairs.

In this Letter, we investigate the fate of the non-Abelian anyons of various even-denominator paired-CF states, which should provide guidance on where they appear most robustly. Specifically, we ask whether they form bound states.  This question might at first seem surprising, as one would expect the Coulomb repulsion to keep them apart. Indeed, if the QPs or QHs were point particles, no bound state would be formed. However, the actual QPs and QHs are large objects with complex internal structure involving density oscillations about the uniform density of the background FQH liquid, and experience has shown that in such cases the two QPs or QHs can often find some configuration in which the gain in correlation energy overcomes the Coulomb barrier, which itself is significantly diminished by the large size of the QP / QH.

 Several classes of trial wave functions have been proposed for the paired CF states. These include the Pfaffian wave function~\cite{Moore91}, certain parton wave functions~\cite{Balram18,Faugno19,Faugno20A,Faugno21}, and a BCS-like wave function~\cite{Sharma21,Sharma24}. 
 The parton wave functions~\cite{Balram18,Faugno19,Faugno20A,Faugno21} are the most convenient for our present purpose, as they  
 provide a reasonably accurate description of the actual incompressible state~\cite{Balram18,Faugno19}, and they 
 can be evaluated for large systems, which is essential for capturing the physics of the bound state, which requires a comparison between two nearby QPs / QHs with far separated ones.  We note that binding of the QPs / QHs of the Jain $n/(2pn\pm 1)$ FQH states~\cite{Gattu25A,Xu25,Wang26A,Li26} and of the QHs of the $5/2$ state~\cite{Gattu25A,Xu25} has been studied previously.

The parton theory of FQHE~\cite{Jain89B} constructs new candidate incompressible states starting from known incompressible states. To accomplish that, one begins by splitting each electron into $k$ fictitious fermionic particles called partons, and places partons of each species in an integer quantum Hall (IQH) state (or, more generally in any known incompressible state). Noting that each parton's position is tied to the position of the parent electron, one ends up with a product wave function, which, after  projection into the lowest Landau level (LLL), takes the form
\begin{equation}
  \Psi_\nu = \LLL \prod_{a=1}^{k}\Phi_{n_{a}},\;\;\nu = \left[\sum_{a=1}^{k} n_a^{-1}\right]^{-1}
\label{eq:parton}
\end{equation}
where $\Phi_{n_{a}}$ is the Slater-determinant wave function of the IQH state at filling factor $n_a$; negative values of $n_a$ are allowed with $\Phi_{-|n|}\equiv [\Phi_{|n|}]^*$. It is assumed that the LLL projection does not alter the topological character of the state. These states are referred to as the $n_1 n_2 ...$ parton states, with the convention $\bar{n}=-n$. They contain all of the Jain CF states, but also many other states including many non-Abelian states -- all states where an integer other than 1 appears more than once have non-Abelian QPs~\cite{Wen91}. The state obtained from the $n_1 n_2 ...$ parton state by particle-hole (PH) transformation will be referred to as the anti-$n_1 n_2 ...$ parton state. We will use the Haldane spherical geometry~\cite{Haldane83,Wu76,Wu77} for all our calculations below (details in Supplemental Material (SM)~\cite{SM}).
Throughout, we ignore LL mixing; therefore, our work does not distinguish between two representations of the $\nu=1/2$ paired state related by PH transformation.

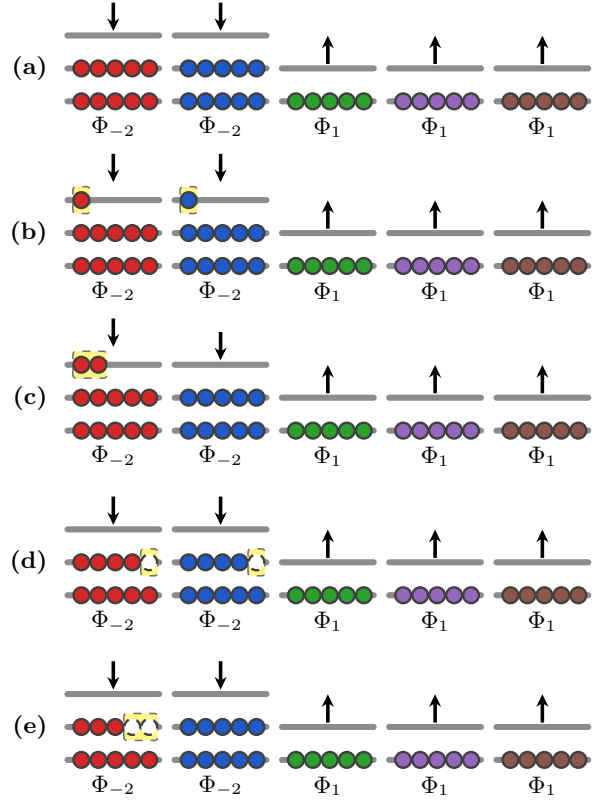
\begin{figure}[t]
  \centering
  \begin{tikzpicture}[scale=0.66]
    \definecolor{pA}{RGB}{214,39,40}
    \definecolor{pB}{RGB}{31,90,200}
    \definecolor{pC}{RGB}{44,160,44}
    \definecolor{pD}{RGB}{148,103,189}
    \definecolor{pE}{RGB}{140,86,75}

    \begin{scope}[shift={(0,0)}]
      \node[font=\small\bfseries,anchor=east] at (-0.25,0.66) {(a)};
      \lamf{0.00}{pA}{$\Phi_{-2}$}{{5,5,0}}{-1}{0}{<-}
      \lamf{2.15}{pB}{$\Phi_{-2}$}{{5,5,0}}{-1}{0}{<-}
      \lamf{4.30}{pC}{$\Phi_{1}$}{{5,0}}{-1}{0}{->}
      \lamf{6.45}{pD}{$\Phi_{1}$}{{5,0}}{-1}{0}{->}
      \lamf{8.60}{pE}{$\Phi_{1}$}{{5,0}}{-1}{0}{->}
    \end{scope}

    \begin{scope}[shift={(0,-3.3)}]
      \node[font=\small\bfseries,anchor=east] at (-0.25,0.66) {(b)};
      \lamf{0.00}{pA}{$\Phi_{-2}$}{{5,5,1}}{2}{0}{<-}
      \lamf{2.15}{pB}{$\Phi_{-2}$}{{5,5,1}}{2}{0}{<-}
      \lamf{4.30}{pC}{$\Phi_{1}$}{{5,0}}{-1}{0}{->}
      \lamf{6.45}{pD}{$\Phi_{1}$}{{5,0}}{-1}{0}{->}
      \lamf{8.60}{pE}{$\Phi_{1}$}{{5,0}}{-1}{0}{->}
    \end{scope}

    \begin{scope}[shift={(0,-6.6)}]
      \node[font=\small\bfseries,anchor=east] at (-0.25,0.66) {(c)};
      \lamf{0.00}{pA}{$\Phi_{-2}$}{{5,5,2}}{2}{0}{<-}
      \lamf{2.15}{pB}{$\Phi_{-2}$}{{5,5,0}}{-1}{0}{<-}
      \lamf{4.30}{pC}{$\Phi_{1}$}{{5,0}}{-1}{0}{->}
      \lamf{6.45}{pD}{$\Phi_{1}$}{{5,0}}{-1}{0}{->}
      \lamf{8.60}{pE}{$\Phi_{1}$}{{5,0}}{-1}{0}{->}
    \end{scope}

    \begin{scope}[shift={(0,-9.9)}]
      \node[font=\small\bfseries,anchor=east] at (-0.25,0.66) {(d)};
      \lamf{0.00}{pA}{$\Phi_{-2}$}{{5,4,0}}{1}{1}{<-}
      \lamf{2.15}{pB}{$\Phi_{-2}$}{{5,4,0}}{1}{1}{<-}
      \lamf{4.30}{pC}{$\Phi_{1}$}{{5,0}}{-1}{0}{->}
      \lamf{6.45}{pD}{$\Phi_{1}$}{{5,0}}{-1}{0}{->}
      \lamf{8.60}{pE}{$\Phi_{1}$}{{5,0}}{-1}{0}{->}
    \end{scope}

    \begin{scope}[shift={(0,-13.2)}]
      \node[font=\small\bfseries,anchor=east] at (-0.25,0.66) {(e)};
      \lamf{0.00}{pA}{$\Phi_{-2}$}{{5,3,0}}{1}{2}{<-}
      \lamf{2.15}{pB}{$\Phi_{-2}$}{{5,5,0}}{-1}{0}{<-}
      \lamf{4.30}{pC}{$\Phi_{1}$}{{5,0}}{-1}{0}{->}
      \lamf{6.45}{pD}{$\Phi_{1}$}{{5,0}}{-1}{0}{->}
      \lamf{8.60}{pE}{$\Phi_{1}$}{{5,0}}{-1}{0}{->}
    \end{scope}
  \end{tikzpicture}
  \caption{Panel (a) schematically depicts the parton ground state $\Psi^{\bar{2}\bar{2}111}$; the arrow on top shows the magnetic field direction. Panels (b) and (d) show two quasiholes (QHs) and two quasiparticles (QPs) (marked by yellow boxes) in the odd-parity channel, and panels (c) and (e) in the even-parity channel. Note that a QP of the parton with filling factor $2$ produces, due to complex conjugation,  a QH of the $\nu=1/2$ state.
  }
  \label{fig:lambda}
\end{figure}

The parton states 
\begingroup\small
\begin{alignat}{2}
\Psi^{221}_{1/2} &= \LLL\,[\Phi_{2}]^{2}\,\Phi_{1},
&\quad \Psi^{\bar{2}\bar{2}111}_{1/2} &= \LLL\,[\conj{\Phi}_{2}]^{2}\,\Phi_{1}^{3}, \label{eq:221} \\[1pt]
\Psi^{22111}_{1/4} &= \LLL\,[\Phi_{2}]^{2}\,\Phi_{1}^{3},
&\quad \Psi^{\bar{2}\bar{2}11111}_{1/4} &= \LLL\,[\conj{\Phi}_{2}]^{2}\,\Phi_{1}^{5}. \label{eq:22111}
\end{alignat}
\endgroup
have been identified in the past as being relevant to the observed states, as listed in Table~\ref{fig:Table}. 
One can add two QPs or two QHs in either different or the same parton species, which we refer to as the odd- and even-parity channels; the total number of CFs in these channels is, respectively, odd and even. See Fig.~\ref{fig:lambda} for a pictorial depiction.

The LLL projection is carried out with the Jain--Kamilla (JK) method~\cite{Jain97A,Jain97}, in its quaternion formulation~\cite{Gattu25}. The $f$-wave ($\ell=3$) 221 state is not amenable to the JK projection, and thus may be projected only for small systems~\cite{Kim19,Wu17}; we will not consider it here. The various states are projected into the LLL as $\Psi^{\bar{2}^21^3}_{1/2}=[\LLL\conj{\Phi}_{2}\Phi_{1}^{2}]^{2}/\Phi_{1}=[\Psi_{2/3}]^2/\Phi_1$, 
$\Psi^{22111}_{1/4} = [\Psi_{2/5}]^2 / \Phi_1$, and $\Psi^{\bar{2}^21^5}_{1/4}=[\Psi_{2/3}]^2\Phi_1$.
The LLL projected states containing two QPs or two QHs are constructed similarly. We place the two QPs or the two QHs in definite relative angular momentum $L_{\rm rel}$, which is an even (odd) integer in the odd (even) parity channel, and evaluate the energy of each state
with the Metropolis--Gibbs--Hastings algorithm; statistical errors are estimated by the jackknife method~\cite{Efron82}. The state in the large $L_{\rm rel}$ limit represents two far separated QPs or QHs.  $\Lrel^{\star}$ denotes the $L_{\rm rel}$ of the lowest energy state. Note that while the energy of two far separated QPs or QHs is independent of which channel they are in, at short distances the energy depends on the channel~\cite{Moller11,Sreejith11,Pu23}.
The binding energy is defined as
\be
\Delta_{2}(\Lrel^{\star}) = E(\Lrel^{\star}) - E(\Lrel\!\to\!\mathrm{max}).
\label{eq:Delta2}
\ee

\begin{figure}[!htb]
  \centering \includegraphics[width=\columnwidth]{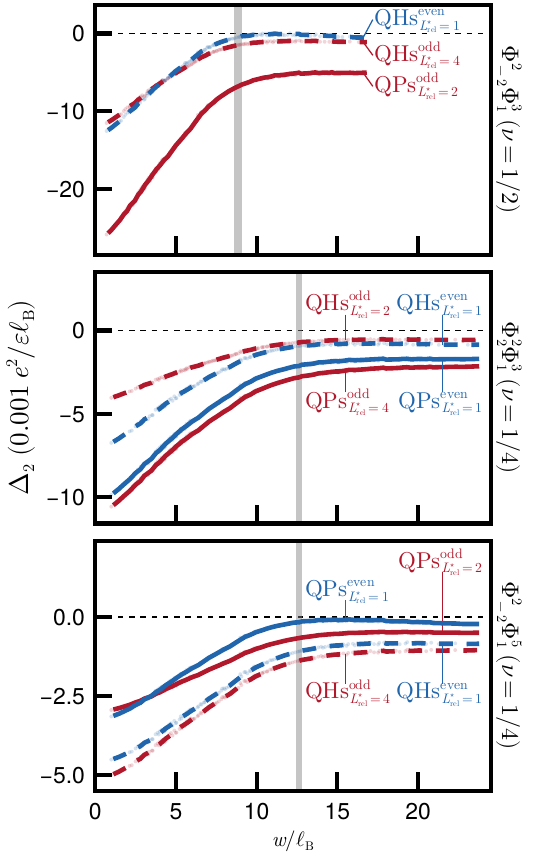}
  \caption{Binding energy $\Delta_{2}$ of two quasiholes (dashed) and two quasiparticles (solid) for $\bar{2}^{2}1^{3}$ at $\nu=1/2$ (top), and $2^{2}1^{3}$ and $\bar{2}^{2}1^{5}$ at $\nu=1/4$ (middle and bottom), as a function of $w/\ell_{B}$, for $N\sim100$ electrons. The width and density dependence approximately collapses onto $w/\ell_{B}$ (Fig.~\ref{fig:wlbdemo}). Blue (red) curves show even (odd) channels; $\Lrel^{\star}$ is indicated. Gray bands mark the experimental onset of the fractional quantum Hall effect, obtained from Figs.~3 and 12 of Ref.~\cite{Sharma24}: $w/\ell_{B}=8.9(2)$ at $\nu=1/2$~\cite{Suen92,Suen94A,Shabani13} and $w/\ell_{B}\approx12.4$--$12.8$ at $\nu=1/4$~\cite{Shabani09,Shabani09A}. Our results are meaningful only to the right of the vertical bands.}
  \label{fig:wlb}
\end{figure}

The results are given in Table~\ref{tab:overview} and SM. For the $\bar{2}^{2}1^{3}$ state at $\nu=5/2$, QHs bind in the odd channel while QPs in both.  In WQWs, the $1/2$ and $1/4$ states are stabilized, believed to be represented by $\bar{2}^{2}1^{3}$ and ${2}^{2}1^{3}$; for completeness, we also consider $\bar{2}^{2}1^{5}$ at $\nu=1/4$. We have considered the QPs and QHs of these states in both even and odd parity channels. We find that the binding energies, obtained with a finite-width interaction derived within a self-consistent local density approximation (see SM), are approximately a function of $w/\ell_B$ (Fig.~\ref{fig:wlb}). In the experimentally relevant region, for the $\bar{2}^{2}1^{3}$ state at $\nu=1/2$, only the two-QP state in the odd channel is bound; the two-QP state in the even channel is unbound and the two-QH state has zero binding energy in both channels within the precision of our calculation. All other two-QP and two-QH states are bound in both channels. In monolayer graphene, in the $n=2$ LL, the QPs bind in the even channel and QHs in the odd channel (Table~\ref{tab:overview} and Fig.~\ref{fig:invN-mlg2}). We are unable to obtain reliable results for the $\nu=1/2$ state in the $n=3$ LL of monolayer graphene~\cite{Kim19} (see SM~\cite{SM}). The binding energies are given in Table~\ref{tab:overview}.

An important implication of our work is that the non-Abelian statistics is often hidden at the lowest temperatures, because the non-Abelian QPs and QHs bind to form Abelian molecules. So far the implicitly accepted picture has been that as we move away from half filling, each additional flux quantum produces two elementary non-Abelian QPs or QHs. Our calculations suggest a different picture, that each additional flux quantum produces a single Abelian molecule in the channel that has the largest binding energy.

In the limit of weak disorder, one expects that the additional QPs / QHs form a Wigner crystal, which is pinned by disorder. The lattice constant of the crystal can distinguish between a crystal of charge-$e/4$ QPs and one of charge-$e/2$ molecules. This may be accessible to scanning tunneling microscopy (STM) experiments which have directly imaged the crystal in the FQH regime~\cite{Tsui24}. The STM experiments may also image the charge density and thus directly measure the charge of the localized object~\cite{Chiu25}.

The binding of non-Abelian anyons into Abelian molecules has obvious implications for several proposals to detect non-Abelian braid statistics. These include: an entropy of $k_B (N_\phi-1)\ln 2$ when $N_\phi$ flux quanta are added to the state creating $2N_\phi$ localized non-Abelian QPs / QHs, which ought to be detectable in thermopower measurements~\cite{Yang09} (see for experiments Refs.~\cite{Melcer2024heat,Boivin26}); in a geometry which has two constrictions forming an island, an absence of interference pattern is predicted when the number of QPs / QHs in the inner island is odd~\cite{DasSarma05,Stern06,Bonderson06,Bonderson06A,Nayak08} (see for experiments Ref.~\cite{Willett09}); proposals for forming qubits using the non-Abelian anyons~\cite{DasSarma05,Nayak08}.

The non-Abelian anyons are not always hidden, however. According to our calculations, the QHs of the $\bar{2}^{2}1^{3}$-parton state, or the QPs of the anti-$\bar{2}^{2}1^{3}$-parton state remain free. We thus identify the $1/2$ state in WQWs as a promising state for the realization and study of non-Abelian braid statistics. A related prediction is that, in the zero temperature limit, the charges bound to an impurity are $e/2$ on one side of $\nu=1/2$ and $e/4$ on the other, and to the extent that the charges measured in interference and shot noise experiments are related to the charges of the bound states~\cite{Gattu25A}, such asymmetry is predicted for those charges as well. Further, a determination of which side yields which charge will identify the topological order of the state (i.e., whether it is $\bar{2}^{2}1^{3}$ or anti-$\bar{2}^{2}1^{3}$); if this approach is confirmed, this might be another relatively simple way of identifying the topological order for certain FQH states. (Within the LLL, $\bar{2}^{2}1^{3}$ and anti-$\bar{2}^{2}1^{3}$ are equally plausible, but it is believed that LL mixing will lift the degeneracy and select one of the two~\cite{Bishara09,Rezayi11,Peterson13,Sodemann13,Pakrouski15,Rezayi17}.) The non-Abelian QPs should also become free at temperatures higher than the binding energy but lower than the excitation gap of the paired state.

All of the above conclusions are obtained within the parton theory, which is reasonably accurate but not exact, and therefore will require further theoretical and experimental scrutiny. Nonetheless, our work reveals an important factor that is relevant in the search for non-Abelian statistics, namely that the non-Abelian QPs may be subsumed into Abelian bound states. 

\prlhead{Acknowledgments.} This work was supported in part by the National Science Foundation under Grant No.~DMR-2404619. M.G. was partially supported by a 2025/2026 Rising Researcher Grant from the Penn State Institute for Computational and Data Sciences (RRID:~SCR\_025154). The authors of this work also recognize the Institute for providing access to computational research infrastructure within the Roar Core Facility (RRID:~SCR\_026424). M.G. thanks the Lodha Theoretical Physics Institute for their hospitality during the summer of 2026. Anthropic's Claude (Fable and Opus models) assisted with the preparation of figures and the coordination and execution of numerical calculations.
\prlhead{Data availability.} The data supporting the findings of this study will be made available upon request.
\label{page:main-end}

\clearpage
\onecolumngrid
\setcounter{page}{1}
\setcounter{figure}{0}
\setcounter{table}{0}
\setcounter{equation}{0}
\renewcommand{\thepage}{S\arabic{page}}
\renewcommand{\thefigure}{S\arabic{figure}}
\renewcommand{\thetable}{S\arabic{table}}
\renewcommand{\theequation}{S\arabic{equation}}
\begin{center}
  {\large\bfseries Supplemental Material for ``Search for free non-Abelian anyons in paired composite-fermion states''}
\end{center}
\twocolumngrid

\setcounter{secnumdepth}{2}

In this Supplemental Material, we provide the technical details of our study of the binding of two quasiparticles (QPs) and of two quasiholes (QHs) of paired composite-fermion (CF) states. In Sec.~\ref{sec:sm-sphere}, we briefly review the spherical geometry used in our calculations. Section~\ref{sec:sm-parton} describes certain properties of the parton wave functions in this geometry, and Sec.~\ref{sec:sm-twoqh} the construction of their two-QH and two-QP states. Section~\ref{sec:sm-mc} gives the details of our evaluation of the binding energy $\Delta_{2}$ of the two-QH and two-QP bound states reported in the main text (see Eq.~\eqref{eq:Delta2} of the main text). 

\section{Spherical geometry}
\label{sec:sm-sphere}
In this section, we briefly review Haldane's spherical geometry~\cite{Haldane83} used in our calculations. In this geometry, electrons move on the surface of a sphere, which has a magnetic monopole of strength $Q$ at its center. The monopole produces a uniform radial magnetic field $\mathbf{B}(\mathbf{r}) = 2Q\phi_{0}\,\mathbf{r}/(4\pi r^{3})$ on the sphere, so the total flux through the surface is $2Q\phi_{0}$. Here, $\phi_{0} = hc / e$ is the flux quantum and $2Q$ is an integer. In terms of the magnetic length $\ell_{B} = \sqrt{\hbar c/eB}$, where $B$ is the strength of the magnetic field on the sphere, the radius of the sphere is $R = \sqrt{Q}\,\ell_{B}$.

The single-particle eigenfunctions are the monopole harmonics~\cite{Wu76,Wu77} $Y_{Q, l, m}(\theta, \phi)$, where $\theta$ and $\phi$ are the polar and azimuthal angles of the electron. They form degenerate shells of fixed orbital angular momentum $l$, which are the Landau levels in this geometry; the orbitals within a shell are labeled by the azimuthal angular momentum $m = -l, -l+1, \dots, l$. 

A convenient representation of the monopole harmonics uses the spinor variables~\cite{Haldane83} $u = \cos(\theta/2)e^{i\phi/2}$ and $v = \sin(\theta/2)e^{-i\phi/2}$, in terms of which~\cite{Jain97A,Jain97,Jain07}
\begin{equation}
  \begin{split}
    &Y_{Q, l, m}(\theta, \phi) = N_{Q, l, m}(-1)^{l-m}v^{Q-m}u^{Q+m} \times \\
    &\sum_{s=0}^{l-m}(-1)^{s}\tbinom{l-Q}{s}\tbinom{l+Q}{l-m-s}(v^{\star}v)^{l-Q-s}(u^{\star}u)^{s}, \\
    &N_{Q, l, m} = \sqrt{\tfrac{(2l+1) (l-m)! (l+m)!}{4\pi (l-Q)! (l+Q)!}}.
  \end{split}
\end{equation}
In the lowest Landau level (LLL) ($l=Q$; in general, the $n^{\rm th}$ Landau level (LL) has $l=Q+n$, with $n=0,1,\dots$), this simplifies to $Y_{Q, Q, m}(\theta, \phi) = N_{Q, Q, m}(-1)^{Q-m}v^{Q-m}u^{Q+m}$.

\section{Parton wave functions in the spherical geometry}
\label{sec:sm-parton}
In this section, we describe the construction of the parton wavefunctions used in this work in the spherical geometry.

The $n_1 n_2 \dots n_k$ parton wave function is given in Eq.~\ref{eq:parton}, where  $\Phi_{n}$ denotes the integer quantum Hall (IQH) state of $N$ electrons filling $n$ LLs. The $n$ filled shells contain $N=n(2Q_n+n)$ orbitals, giving
\be
2Q_n=\frac{N}{n}-n.
\label{eq:sm2-iqh-flux}
\ee
In general, a state at $\nu$ occurs at $2Q=\nu^{-1}N-S$ where $S$ is called the shift. The monopole strength and the shift for the $n_1 n_2 \dots n_k$ parton wave function are given by
\be
2Q=\sum_{a=1}^k 2Q_a,\;\; S=\sum_{a=1}^k n_a.
\ee
The parton states studied in this work involve partons at $|\nu|=2$ and $\nu=1$. Therefore, the factors entering Eq.~\eqref{eq:parton} of the main text are $\Phi_{2}$ or its complex conjugate and $\Phi_{1}$. 

Using the spinors defined in Sec.~\ref{sec:sm-sphere}, the filled-LLL Slater determinant $\Phi_{1}$ can be written as
\be
\begin{split}
\Phi_1 &\propto
\begin{vmatrix}
u_1^{N-1} & \cdots & u_N^{N-1} \\
u_1^{N-2}v_1 & \cdots & u_N^{N-2}v_N \\
\vdots & \ddots & \vdots \\
v_1^{N-1} & \cdots & v_N^{N-1}
\end{vmatrix} \\
&=\prod_{i<j}(u_i v_j-u_j v_i).
\end{split}
\label{eq:sm2-phi1}
\ee

For even $N$, $\Phi_{2}$ fills the shells $l=Q$ and $l=Q+1$, containing $2Q+1$ and $2Q+3$ orbitals, respectively. Its Slater determinant is
\be
\Phi_2 \propto
\begin{vmatrix}
Y_{Q,Q,-Q}(\Omega_1) & \cdots & Y_{Q,Q,-Q}(\Omega_N) \\
\vdots & \ddots & \vdots \\
Y_{Q,Q,Q}(\Omega_1) & \cdots & Y_{Q,Q,Q}(\Omega_N) \\
Y_{Q,Q+1,-Q-1}(\Omega_1) & \cdots & Y_{Q,Q+1,-Q-1}(\Omega_N) \\
\vdots & \ddots & \vdots \\
Y_{Q,Q+1,Q+1}(\Omega_1) & \cdots & Y_{Q,Q+1,Q+1}(\Omega_N)
\end{vmatrix},
\label{eq:sm2-phi2}
\ee
where $\Omega_i=(\theta_i,\phi_i)$ is shorthand for the position of the $i^{\rm th}$ electron. Using these, the parton states of Eq.~\eqref{eq:parton} of the main text may be constructed.

\section{Two-quasihole and two-quasiparticle states}
\label{sec:sm-twoqh}

In this section, we detail the construction of the two-QH and two-QP wavefunctions in the odd and even channels described in the main text, at a given 
relative angular momentum $L_{\rm rel}$. For simplicity, we focus on the $\bar{2}^{2}1^{3}$ parton phase. The construction for other states is analogous.

\textit{Odd channel.} In the odd channel, two QHs or two QPs are created by placing one QH or QP in each $\nu=-2$ parton factor; the electron number $N$ is odd.

The lowest-energy QP states are created by removing a parton from the second LL ($n_{\rm LL}=1$) in each factor. Labeling the two holes by their azimuthal angular momenta $(m_1,m_2)$ before conjugation gives the product wavefunction
\be
\Psi_{1/2}^{\mathrm{QP,odd};m_1,m_2}
=\LLL\,\conj{\Phi}_2^{-;m_1}\conj{\Phi}_2^{-;m_2}\Phi_1^3.
\label{eq:sm3-qp-odd}
\ee
Each $\Phi_2^{-;m}$ is an $N$-particle determinant at $Q=(N-3)/4$:
\be
\Phi_2^{-;m}\propto
\begin{vmatrix}
Y_{Q,Q,-Q}(\Omega_1) & \cdots & Y_{Q,Q,-Q}(\Omega_N) \\
\vdots & \ddots & \vdots \\
Y_{Q,Q,Q}(\Omega_1) & \cdots & Y_{Q,Q,Q}(\Omega_N) \\
Y_{Q,Q+1,-Q-1}(\Omega_1) & \cdots & Y_{Q,Q+1,-Q-1}(\Omega_N) \\
\vdots & \ddots & \vdots \\
Y_{Q,Q+1,-m-1}(\Omega_1) & \cdots & Y_{Q,Q+1,-m-1}(\Omega_N) \\
Y_{Q,Q+1,-m+1}(\Omega_1) & \cdots & Y_{Q,Q+1,-m+1}(\Omega_N) \\
\vdots & \ddots & \vdots \\
Y_{Q,Q+1,Q+1}(\Omega_1) & \cdots & Y_{Q,Q+1,Q+1}(\Omega_N)
\end{vmatrix},
\label{eq:sm3-removed-determinant}
\ee
The hole has azimuthal angular momentum $m$ before conjugation.

Similarly, the lowest-energy QH states are created by adding a parton to the third LL ($n_{\rm LL}=2$) in each factor:
\be
\Psi_{1/2}^{\mathrm{QH,odd};m_1,m_2}
=\LLL\,\conj{\Phi}_2^{+;m_1}\conj{\Phi}_2^{+;m_2}\Phi_1^3.
\label{eq:sm2-top}
\ee
Each $\Phi_2^{+;m}$ is an $N$-particle determinant at $Q=(N-5)/4$:
\be
\Phi_2^{+;m}\propto
\begin{vmatrix}
Y_{Q,Q,-Q}(\Omega_1) & \cdots & Y_{Q,Q,-Q}(\Omega_N) \\
\vdots & \ddots & \vdots \\
Y_{Q,Q,Q}(\Omega_1) & \cdots & Y_{Q,Q,Q}(\Omega_N) \\
Y_{Q,Q+1,-Q-1}(\Omega_1) & \cdots & Y_{Q,Q+1,-Q-1}(\Omega_N) \\
\vdots & \ddots & \vdots \\
Y_{Q,Q+1,Q+1}(\Omega_1) & \cdots & Y_{Q,Q+1,Q+1}(\Omega_N) \\
Y_{Q,Q+2,m}(\Omega_1) & \cdots & Y_{Q,Q+2,m}(\Omega_N)
\end{vmatrix}.
\label{eq:sm3-added-determinant}
\ee

We obtain states at a given total angular momentum $(L,M)$ by combining the different Slater-determinant product states with Clebsch--Gordon coefficients:
\be
\begin{split}
\Psi_{\nu}^{X,\mathrm{odd};L,M}\propto
\sum_{m_1}
&\overlap{l_X,m_1;l_X,\mu-m_1}{L,\mu}\\
&\times\Psi_{\nu}^{X,\mathrm{odd};m_1,\mu-m_1},
\end{split}
\label{eq:sm3-odd-coupling}
\ee
where $X=\mathrm{QH}$ or $\mathrm{QP}$ and $l_X$ is the angular momentum of an individual QH or QP. Here, $\mu=-M$ accounts for the complex conjugation of the parton factors. We define the relative angular momentum as
\be
\Lrel \equiv 2l_X-L.
\label{eq:sm2-Lrel}
\ee
Interchanging $m_1$ and $m_2$ leaves the product wavefunction unchanged, so only the even values $\Lrel=0,2,4,\dots$ are allowed in the odd channel.

\textit{Even channel.} In the even channel, both QHs or both QPs reside in the same $\nu=-2$ parton factor; the electron number $N$ is even.

The lowest-energy QP states are created by removing two partons from the second LL ($n_{\rm LL}=1$) of one factor. Labeling the two holes by their azimuthal angular momenta $(m_1,m_2)$ before conjugation gives the product wavefunction
\be
\Psi_{1/2}^{\mathrm{QP,even};m_1,m_2}
=\LLL\,\conj{\Phi}_2^{-;m_1,m_2}\conj{\Phi}_2\Phi_1^3.
\label{eq:sm3-qp-even}
\ee
The determinant $\Phi_2^{-;m_1,m_2}$, at $Q=(N-2)/4$, can be written as
\be
\begin{gathered}
\Phi_2^{-;m_1,m_2}\propto\\
\begin{vmatrix}
Y_{Q,Q,-Q}(\Omega_1) & \cdots & Y_{Q,Q,-Q}(\Omega_N) \\
\vdots & \ddots & \vdots \\
Y_{Q,Q,Q}(\Omega_1) & \cdots & Y_{Q,Q,Q}(\Omega_N) \\
Y_{Q,Q+1,-Q-1}(\Omega_1) & \cdots & Y_{Q,Q+1,-Q-1}(\Omega_N) \\
\vdots & \ddots & \vdots \\
Y_{Q,Q+1,-m_2-1}(\Omega_1) & \cdots & Y_{Q,Q+1,-m_2-1}(\Omega_N) \\
Y_{Q,Q+1,-m_2+1}(\Omega_1) & \cdots & Y_{Q,Q+1,-m_2+1}(\Omega_N) \\
\vdots & \ddots & \vdots \\
Y_{Q,Q+1,-m_1-1}(\Omega_1) & \cdots & Y_{Q,Q+1,-m_1-1}(\Omega_N) \\
Y_{Q,Q+1,-m_1+1}(\Omega_1) & \cdots & Y_{Q,Q+1,-m_1+1}(\Omega_N) \\
\vdots & \ddots & \vdots \\
Y_{Q,Q+1,Q+1}(\Omega_1) & \cdots & Y_{Q,Q+1,Q+1}(\Omega_N)
\end{vmatrix}.
\end{gathered}
\label{eq:sm3-even-removed-determinant}
\ee
The other $\Phi_2$ factor has two filled shells at $Q=(N-4)/4$.

Similarly, the lowest-energy QH states are created by adding two partons to the third LL ($n_{\rm LL}=2$) of one factor:
\be
\Psi_{1/2}^{\mathrm{QH,even};m_1,m_2}
=\LLL\,\conj{\Phi}_2^{+;m_1,m_2}\conj{\Phi}_2\Phi_1^3.
\label{eq:sm2-nontop}
\ee
The determinant $\Phi_2^{+;m_1,m_2}$, at $Q=(N-6)/4$, can be written as
\be
\Phi_2^{+;m_1,m_2}\propto
\begin{vmatrix}
Y_{Q,Q,-Q}(\Omega_1) & \cdots & Y_{Q,Q,-Q}(\Omega_N) \\
\vdots & \ddots & \vdots \\
Y_{Q,Q,Q}(\Omega_1) & \cdots & Y_{Q,Q,Q}(\Omega_N) \\
Y_{Q,Q+1,-Q-1}(\Omega_1) & \cdots & Y_{Q,Q+1,-Q-1}(\Omega_N) \\
\vdots & \ddots & \vdots \\
Y_{Q,Q+1,Q+1}(\Omega_1) & \cdots & Y_{Q,Q+1,Q+1}(\Omega_N) \\
Y_{Q,Q+2,m_1}(\Omega_1) & \cdots & Y_{Q,Q+2,m_1}(\Omega_N) \\
Y_{Q,Q+2,m_2}(\Omega_1) & \cdots & Y_{Q,Q+2,m_2}(\Omega_N)
\end{vmatrix}.
\label{eq:sm3-even-added-determinant}
\ee
The other $\Phi_2$ factor is again at $Q=(N-4)/4$.

As in the odd channel, we obtain states at a given total angular momentum $(L,M)$ by combining the different Slater-determinant product states with Clebsch--Gordon coefficients:
\be
\begin{split}
\Psi_{\nu}^{X,\mathrm{even};L,M}\propto
\sum_{\substack{m_1<m_2\\m_1+m_2=\mu}}
&\overlap{l_X,m_1;l_X,m_2}{L,\mu}\\
&\times\Psi_{\nu}^{X,\mathrm{even};m_1,m_2},
\end{split}
\label{eq:sm3-pair-determinants}
\ee
where $\mu$ is defined as in the odd channel. Since the two added partons or two holes are identical fermions in the same LL, only the odd values $\Lrel=1,3,5,\dots$ are allowed in the even channel.

For the rotationally invariant interactions considered here, different $M$ states at fixed $L$ are degenerate, and we use the maximum-weight state $M=L$ throughout.

\begin{figure*}[t]
\centering
  \includegraphics[width=0.94\textwidth,height=0.72\textheight,keepaspectratio]{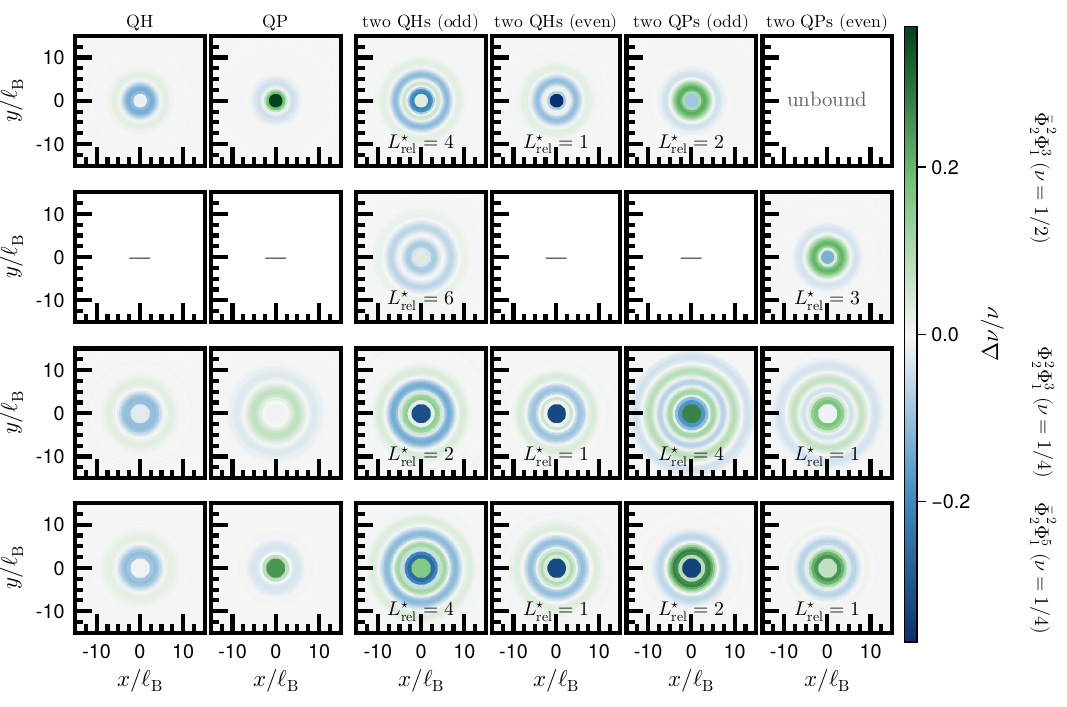}
  \caption{Density profiles of quasiholes (QHs), quasiparticles (QPs), and their bound states for the three parton phases labeled on the right, for $N\simeq120$, shown as the local filling-factor deviation $\Delta\nu(x,y)/\nu$. The first two rows correspond to $\bar{2}^{2}1^{3}$. First two columns: a single QH and a single QP pinned at the origin. Last four columns: the lowest-energy two-QH and two-QP states of the odd and even channels of Table~\ref{tab:lrel}, pinned again at the origin; the relative angular momentum $\Lrel^{\star}$ is indicated in each panel. The profiles are densities of lowest-Landau-level (LLL) trial states, without the form-factor conversion to higher Landau levels. Dashes denote profiles not shown. The even two-QP channel of $\bar{2}^{2}1^{3}$ is unbound for the LLL interaction and is not shown.}
  \label{fig:density}
\end{figure*}
\begin{figure*}[t]
\centering
  \includegraphics[width=0.94\textwidth,height=0.40\textheight,keepaspectratio]{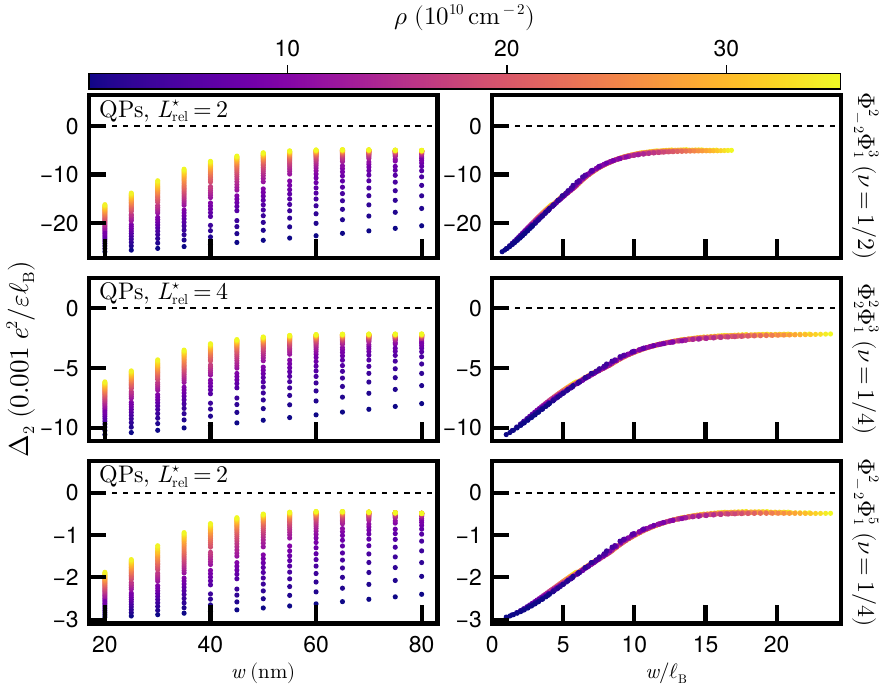}
  \caption{Collapse of the well-width and density dependence of $\Delta_{2}$ onto the single variable $w/\ell_{B}$, shown for the odd two-quasiparticle channel of each parton phase (top to bottom, annotated at the right edge), for $N \sim 100$ electrons. $\Delta_{2}$ is defined as in Fig.~\ref{fig:invN}. The left column shows $\Delta_{2}$ as a function of the well width $w$, with the color of each point depicting the density $\rho$; the right column shows the same data as a function of $w/\ell_{B}$.}
  \label{fig:wlbdemo}
\end{figure*}

\section{Effective interactions}
\label{sec:sm-interactions}
\label{sec:sm-mc}

In this section, we describe the three settings in which we examine even-denominator FQH states: the LLL in wide quantum wells, higher LLs in the ideal two-dimensional limit (narrow wells), and higher LLs of monolayer graphene. In each setting, we neglect LL mixing and transverse subband mixing, and map the physics onto the ideal two-dimensional LLL through an effective interaction.

\subsection{Wide quantum wells}
For a quantum well of width $w$ and density $\rho$, we obtain the transverse density profile $\varrho(z)$, normalized as $\int dz\,\varrho(z)=1$, in a zero-field local-density approximation~\cite{Ortalano97,Rother09}, and use the effective two-dimensional interaction
\begin{equation}
V(r)=\frac{e^{2}}{\epsilon}\int dz\,dz'\,\frac{\varrho(z)\,\varrho(z')}{\sqrt{r^{2}+(z-z')^{2}}}.
\end{equation}

\subsection{Higher Landau levels of GaAs}
The Coulomb interaction in the $n^{\rm th}$ LL is specified by its Haldane pseudopotentials. On the sphere we take the orbitals of the $n^{\rm th}$ LL to have angular momentum $l=Q$, i.e., monopole strength $Q-n$, so that the $n^{\rm th}$ LL has the same single-particle Hilbert space as the LLL at monopole strength $Q$. The pseudopotential for a pair with relative angular momentum $L$ (total angular momentum $2Q-L$) is then
\begin{equation}
\begin{split}
V^{(n)}_{L}={}&\frac{e^{2}(2Q+1)^{2}}{\epsilon R}(-1)^{L}\\
&\times\sum_{k=0}^{2Q}
\begin{Bmatrix} 2Q-L & Q & Q\\ k & Q & Q\end{Bmatrix}
\left[F_{n}(k)\right]^{2},
\end{split}
\label{eq:sm-VLn}
\end{equation}
with the form factor
\begin{equation}
F_{n}(k)=\begin{pmatrix} Q & Q & k\\ Q-n & -(Q-n) & 0\end{pmatrix}.
\label{eq:sm-Fn}
\end{equation}
Here, the array in braces is a Wigner $6j$ symbol and the array in parentheses is a Wigner $3j$ symbol~\cite{Varshalovich88}. We map this problem onto the LLL following Ref.~\cite{Yutushui25}.

\subsection{Higher Landau levels of monolayer graphene}
In the $n^{\rm th}$ ($n\geq1$) LL of monolayer graphene (MLG), the single-particle states are two-component spinors built from LL indices $n-1$ and $n$. Near the $K$ valley, at low-energies, the electron dynamics is governed by the Dirac Hamiltonian~\cite{CastroNeto09, Goerbig11}
\begin{equation}
H=v_{F}\left(\sigma_{x}\pi_{x}+\sigma_{y}\pi_{y}\right)=\frac{\sqrt{2}\hbar v_{F}}{\ell_{B}}\begin{pmatrix} 0 & a\\ a^{\dagger} & 0\end{pmatrix},
\end{equation}
where $\boldsymbol{\pi}=\mathbf{p}+e\mathbf{A}/c$, $v_{F}$ is the Fermi velocity, and $a=\ell_{B}(\pi_{x}-i\pi_{y})/(\sqrt{2}\hbar)$ satisfies $[a,a^{\dagger}]=1$. The eigenstates are
\begin{equation}
\psi_{\pm n,m}=\frac{1}{\sqrt{2}}\begin{pmatrix} |n-1,m\rangle\\ \pm|n,m\rangle\end{pmatrix},\qquad E_{\pm n}=\pm\frac{\sqrt{2}\hbar v_{F}}{\ell_{B}}\sqrt{n},
\end{equation}
for $n\geq1$, and $\psi_{0,m}=(0,\,|0,m\rangle)^{T}$ with $E_{0}=0$, where $|n,m\rangle$ is the $m^{\rm th}$ orbital of the $n^{\rm th}$ LL of a parabolic-band electron. Following Ref.~\cite{Balram15D}, in the spherical geometry, we build the spinors from a pair of orbitals with the same angular momentum $l=Q$, and thus with different monopole strengths, $Q-n+1$ and $Q-n$. The pseudopotentials are then given by Eq.~\eqref{eq:sm-VLn} with the form factor
\begin{equation}
F^{\rm MLG}_{n}(k)=\frac{1}{2}\left[F_{n}(k)-F_{n-1}(k)\right],
\end{equation}
where $F_{n}(k)$ is defined in Eq.~\eqref{eq:sm-Fn}. The problem is mapped onto the LLL as before, i.e., following Ref.~\cite{Yutushui25}.

With these interactions, the energies of the two-QH and two-QP states are evaluated with the Metropolis-Hastings-Gibbs algorithm, with the LLL projection carried out by the Jain-Kamilla method~\cite{Jain97A,Jain97,Gattu25}; statistical errors are estimated by the jackknife method~\cite{Efron82}.

\section{Results}
\label{sec:sm-results}
\begin{table}[!hb]
  \caption{Relative angular momentum $\Lrel^{\star}$ of the lowest-energy two-quasihole (QH) and two-quasiparticle (QP) states for the three parton phases ($N\simeq70$). A dash denotes an unbound pair ($\Delta_{2}\ge0$ within the error); a star marks an entry that cannot be resolved because the Monte Carlo errors are too large. For each pair, the first row is the odd (even $\Lrel$) and the second the even (odd $\Lrel$) channel. We show in bold the channel with the lower energy; when the energies are too close to tell we mark neither. The values are unchanged at finite well widths in GaAs, with one exception: for the even two-QP state of $\bar{2}^{2}1^{5}$, the energies at $\Lrel=1$ and $\Lrel=5$ are degenerate within errors, and the minimum moves from $\Lrel=1$ to $\Lrel=5$ for $w/\ell_{B}\gtrsim5$.}
  \label{tab:lrel}
  \footnotesize\setlength{\tabcolsep}{3.5pt}
  \begin{ruledtabular}
  \begin{tabular}{llccccc}
   & & \multicolumn{3}{c}{Graphene} & \multicolumn{2}{c}{GaAs}\\
  \cline{3-5}\cline{6-7}
  phase & pair & $n=0$ & $n=1$ & $n=2$ & $n=0$ & $n=1$\\
  \hline
  \multirow{4}{*}{$\bar{2}^{2}1^{3}$} & QHs & 4 & 4 & \textbf{6} & 4 & \textbf{6}\\
   & QHs & 1 & 1 & -- & 1 & --\\
   & QPs & \textbf{2} & \textbf{2} & -- & \textbf{2} & 2\\
   & QPs & -- & 1 & \textbf{3} & -- & 3\\
  \hline
  \multirow{4}{*}{$2^{2}1^{3}$} & QHs & 2 & 2 & 2 & 2 & 2\\
   & QHs & \textbf{1} & \textbf{1} & \textbf{1} & \textbf{1} & \textbf{1}\\
   & QPs & \textbf{4} & 4 & 4 & \textbf{4} & 4\\
   & QPs & 1 & 1 & \textbf{7} & 1 & 1\\
  \hline
  \multirow{4}{*}{$\bar{2}^{2}1^{5}$} & QHs & 4 & \textbf{4} & 4 & 4 & \textbf{4}\\
   & QHs & 1 & 1 & 1 & 1 & 1\\
   & QPs & 2 & 2 & $*$ & 2 & 2\\
   & QPs & 1 & 1 & \textbf{1} & 1 & 1\\
  \end{tabular}
  \end{ruledtabular}
\end{table}

In this section, we first show the energy of the two-QH and two-QP states of the three parton phases as a function of their relative angular momentum in both the odd and even channels. We then show the binding energy of the bound state (when it exists) as a function of system size $N$, defined with respect to the state of maximum relative angular momentum, wherein the two QHs or QPs are far separated.

\begin{figure*}[tp]
  \centering
  \includegraphics[width=0.94\textwidth,height=0.72\textheight,keepaspectratio]{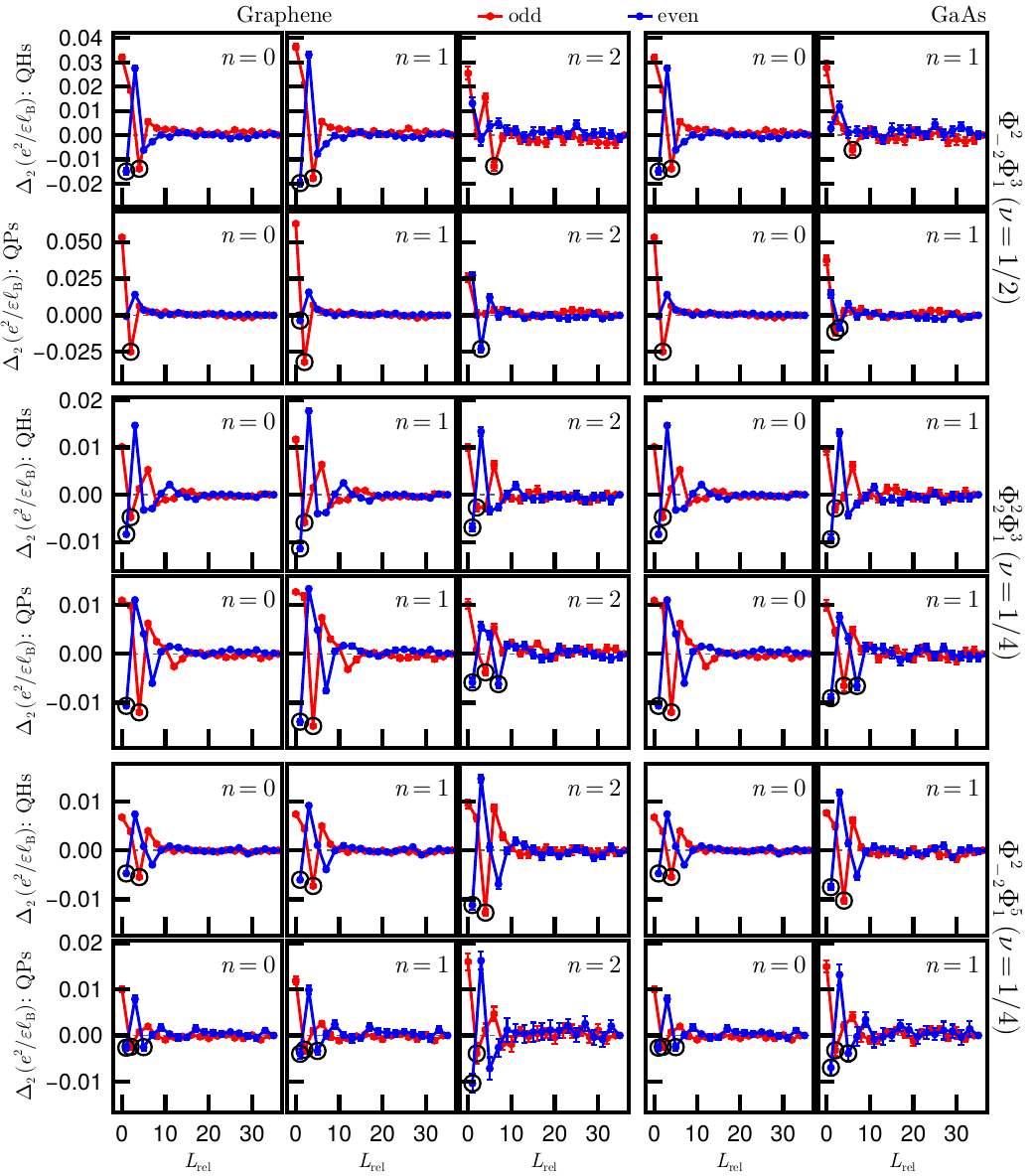}
  \caption{Energy of the two-quasihole and two-quasiparticle states of the three parton phases as a function of their relative angular momentum $\Lrel$, for $N\simeq70$ electrons: $\Phi_{-2}^{2}\Phi_{1}^{3}$ at $\nu=1/2$ (top block), $\Phi_{2}^{2}\Phi_{1}^{3}$ (middle) and $\Phi_{-2}^{2}\Phi_{1}^{5}$ (bottom) at $\nu=1/4$, as annotated at the right edge of each block. The energy is defined with respect to that of the state of maximum relative angular momentum, which in the thermodynamic limit corresponds to two quasiholes (quasiparticles) asymptotically far from each other. Left group: Landau levels $n=0,1,2$ of monolayer graphene; right group: Landau levels $n=0,1$ of GaAs (the $n=0$ panels of the two groups coincide, both being the lowest Landau level). In each panel the odd (even $\Lrel$) and even (odd $\Lrel$) channels are shown together; an open circle marks the lowest-energy bound state of each channel, and where two minima are nearly degenerate, within a channel or between the two channels, both are circled. Error bars are one standard deviation over the Monte Carlo chains.}
  \label{fig:coulomb-disp}\label{fig:quarter-2b2b11111}\label{fig:quarter-22111}
\end{figure*}

In Fig.~\ref{fig:coulomb-disp}, we plot the energy $\Delta_{2}$ of the two-QH and two-QP states of all three parton phases as a function of their relative angular momentum $\Lrel$, for the Coulomb interaction projected into the $n=0$ and $n=1$ LLs of GaAs (corresponding to $\nu=1/2$ and $\nu=5/2$) and the $n=2$ LL of monolayer graphene (corresponding, for one spin-valley copy, to $\nu=13/2$). The relative angular momenta of the bound states, if and when they form, are listed in Table~\ref{tab:lrel}. As mentioned earlier, even $\Lrel$ corresponds to the odd channel and odd $\Lrel$ to the even channel of the two-QH (two-QP) states.

For the $n=0$ LL of GaAs, we show the zero-width limit, since this is simpler than showing the results for each value of $(w,\rho)$. For $N\simeq70$, whenever a bound state exists, its relative angular momentum $\Lrel^{\star}$ is the same as in the zero-width limit for all values of $w=20$--$80$\,nm and $\rho=(1$--$35)\times10^{10}$\,cm$^{-2}$, with one exception: for the even two-QP state of $\bar{2}^{2}1^{5}$, the energies at $\Lrel=1$ and $\Lrel=5$ are degenerate within errors. The even two-QP state of $\bar{2}^{2}1^{3}$ is unbound for all $(w,\rho)$.

For completeness, Fig.~\ref{fig:coulomb-disp} also shows the $n=0$ and $n=1$ LLs of graphene. The $n=0$ LL of graphene is identical to the $n=0$ LL of GaAs, since its single-particle states, $\psi_{0,m}=(0,\,|0,m\rangle)^{T}$, are LLL orbitals. In the $n=1$ LL of graphene, the bound states have the same $\Lrel^{\star}$ as in the LLL, except that the even two-QP state of $\bar{2}^{2}1^{3}$ is bound at $\Lrel^{\star}=1$ for $N\simeq70$, but not significantly at $N=120$. We have not been able to obtain reliable results for the $n=3$ LL of graphene, due to the strongly oscillatory nature of the effective interaction and the resulting difficulty in evaluating the energies by Monte Carlo integration.

\begin{figure*}[tp]
\begin{minipage}[t]{0.48\textwidth}
  \centering
  \includegraphics[width=0.93\linewidth,keepaspectratio]{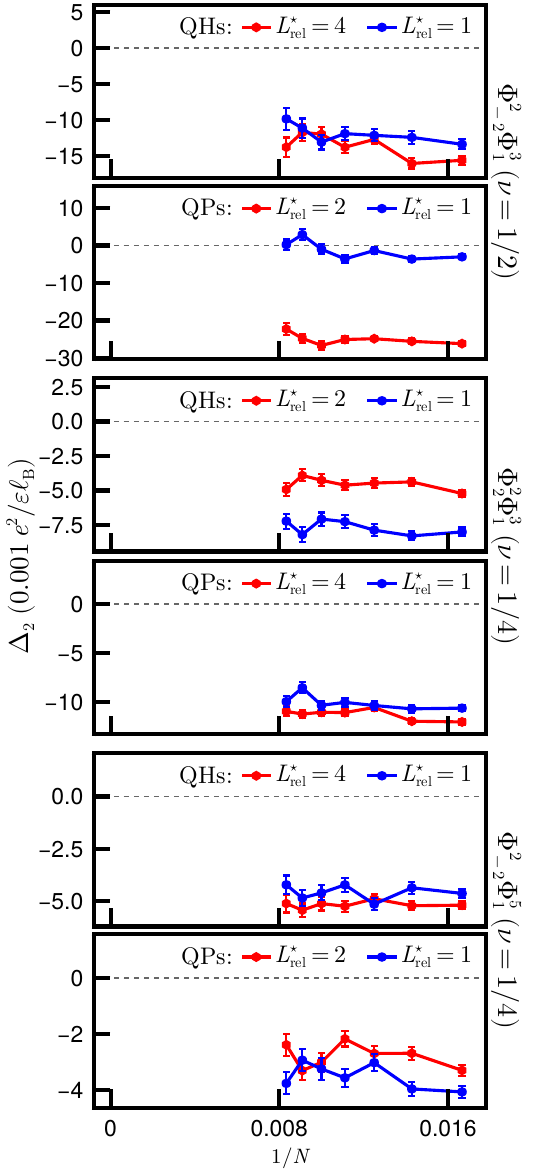}
  \caption{Energy $\Delta_{2}$ of the bound two-quasihole and two-quasiparticle states, at the fixed relative angular momenta $\Lrel^{\star}$ given in the legends, as a function of $1/N$ ($N=60$--$120$) for the lowest-Landau-level Coulomb interaction: $\bar{2}^{2}1^{3}$ at $\nu=1/2$ (top pair of panels), $2^{2}1^{3}$ (middle pair) and $\bar{2}^{2}1^{5}$ (bottom pair) at $\nu=1/4$, as annotated at the right edge. The energy is measured relative to the state of maximum $\Lrel$, Eq.~\eqref{eq:Delta2} of the main text.}
  \label{fig:invN}\label{fig:invN-mlg0}\label{fig:invN-gaas0}
\end{minipage}\hfill
\begin{minipage}[t]{0.48\textwidth}
  \centering
  \includegraphics[width=0.93\linewidth,keepaspectratio]{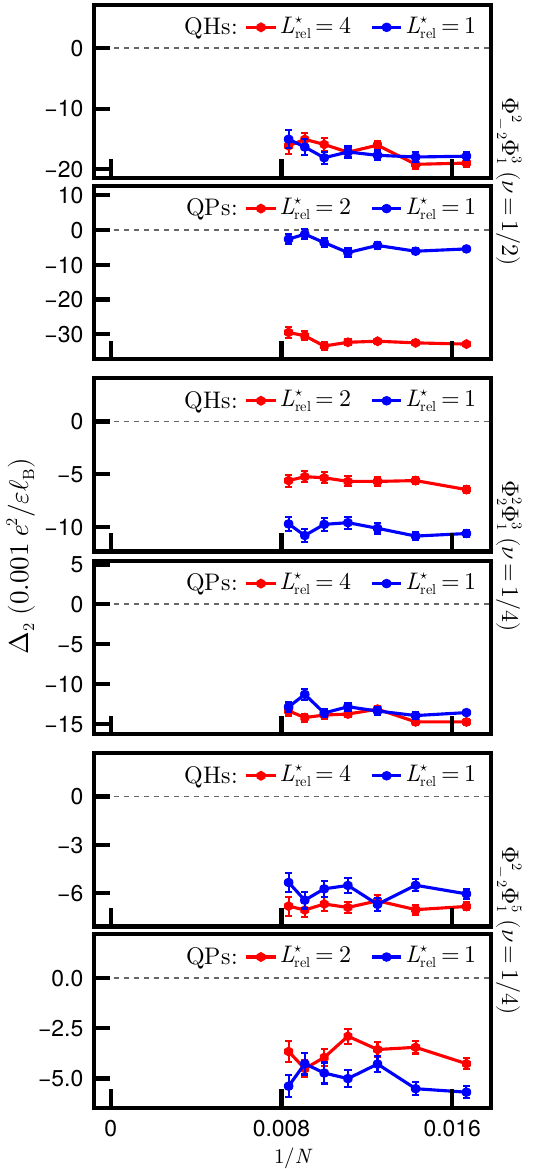}
  \caption{Same as Fig.~\ref{fig:invN}, for the Coulomb interaction projected into the $n=1$ Landau level of monolayer graphene.}
  \label{fig:invN-mlg1}
\end{minipage}
\end{figure*}

\begin{figure*}[tp]
\begin{minipage}[t]{0.48\textwidth}
  \centering
  \includegraphics[width=0.93\linewidth,keepaspectratio]{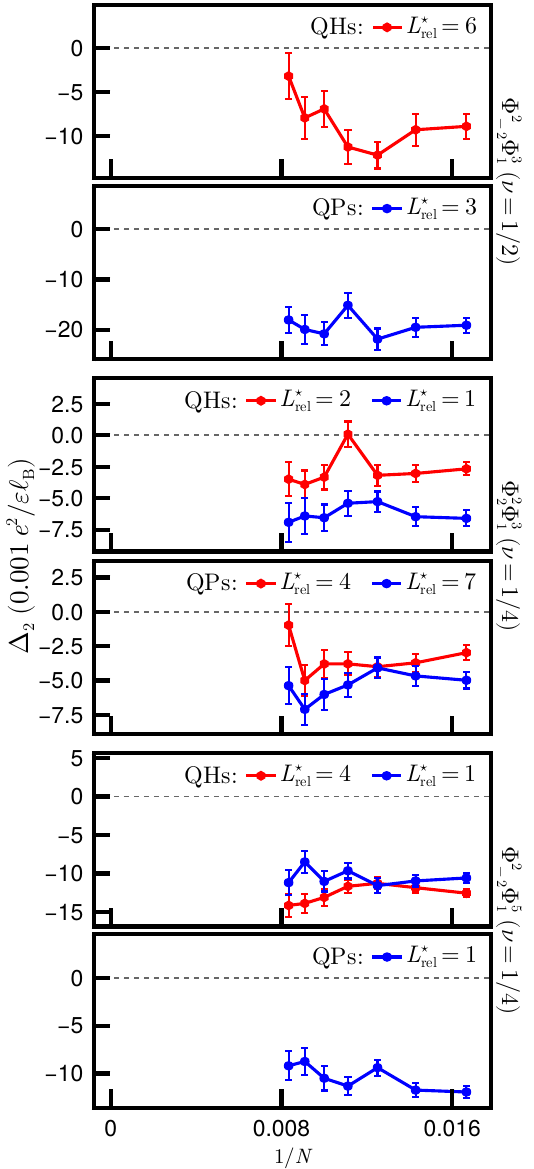}
  \caption{Same as Fig.~\ref{fig:invN}, for the Coulomb interaction projected into the $n=2$ Landau level of monolayer graphene.}
  \label{fig:invN-mlg2}
\end{minipage}\hfill
\begin{minipage}[t]{0.48\textwidth}
  \centering
  \includegraphics[width=0.93\linewidth,keepaspectratio]{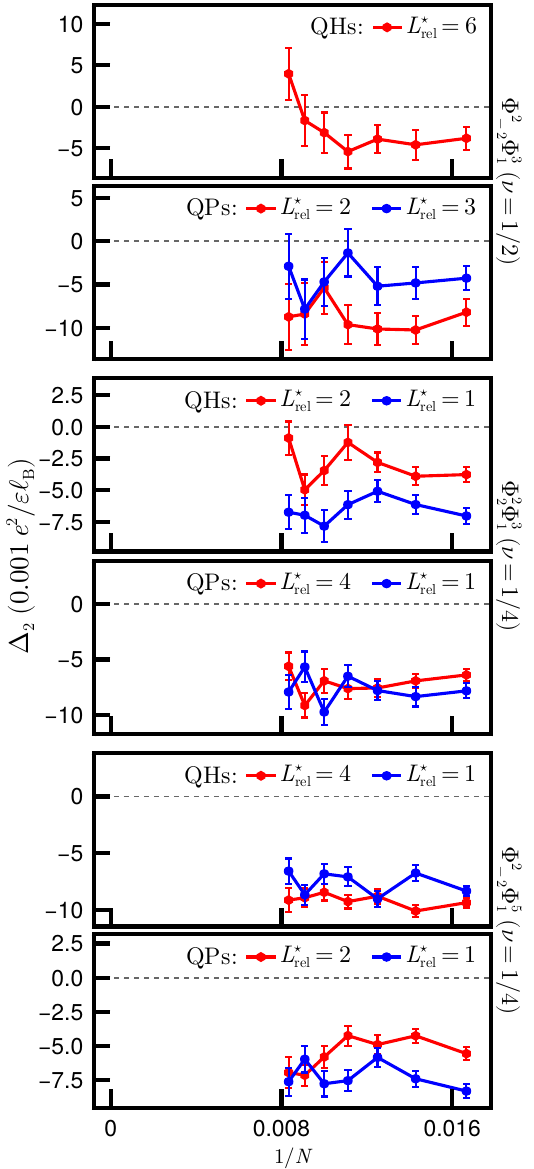}
  \caption{Same as Fig.~\ref{fig:invN}, for the Coulomb interaction projected into the $n=1$ Landau level of GaAs.}
  \label{fig:invN-gaas1}
\end{minipage}
\end{figure*}

Figures~\ref{fig:invN}--\ref{fig:invN-gaas1} show $\Delta_{2}$ of the bound two-QH and two-QP states as a function of $1/N$ ($N=60$--$120$), for the Coulomb interaction projected into the LLL, the $n=1$ and $2$ Landau levels of monolayer graphene, and the $n=1$ Landau level of GaAs. The relative angular momentum of the bound state in each Landau level is given in the legends; channels that are unbound or too noisy to resolve in a given Landau level are omitted.

\begin{thebibliography}{82}%
\makeatletter
\providecommand \@ifxundefined [1]{%
 \@ifx{#1\undefined}
}%
\providecommand \@ifnum [1]{%
 \ifnum #1\expandafter \@firstoftwo
 \else \expandafter \@secondoftwo
 \fi
}%
\providecommand \@ifx [1]{%
 \ifx #1\expandafter \@firstoftwo
 \else \expandafter \@secondoftwo
 \fi
}%
\providecommand \natexlab [1]{#1}%
\providecommand \enquote  [1]{``#1''}%
\providecommand \bibnamefont  [1]{#1}%
\providecommand \bibfnamefont [1]{#1}%
\providecommand \citenamefont [1]{#1}%
\providecommand \href@noop [0]{\@secondoftwo}%
\providecommand \href [0]{\begingroup \@sanitize@url \@href}%
\providecommand \@href[1]{\@@startlink{#1}\@@href}%
\providecommand \@@href[1]{\endgroup#1\@@endlink}%
\providecommand \@sanitize@url [0]{\catcode `\\12\catcode `\$12\catcode
  `\&12\catcode `\#12\catcode `\^12\catcode `\_12\catcode `\%12\relax}%
\providecommand \@@startlink[1]{}%
\providecommand \@@endlink[0]{}%
\providecommand \url  [0]{\begingroup\@sanitize@url \@url }%
\providecommand \@url [1]{\endgroup\@href {#1}{\urlprefix }}%
\providecommand \urlprefix  [0]{URL }%
\providecommand \Eprint [0]{\href }%
\providecommand \doibase [0]{https://doi.org/}%
\providecommand \selectlanguage [0]{\@gobble}%
\providecommand \bibinfo  [0]{\@secondoftwo}%
\providecommand \bibfield  [0]{\@secondoftwo}%
\providecommand \translation [1]{[#1]}%
\providecommand \BibitemOpen [0]{}%
\providecommand \bibitemStop [0]{}%
\providecommand \bibitemNoStop [0]{.\EOS\space}%
\providecommand \EOS [0]{\spacefactor3000\relax}%
\providecommand \BibitemShut  [1]{\csname bibitem#1\endcsname}%
\let\auto@bib@innerbib\@empty
\bibitem [{\citenamefont {Willett}\ \emph {et~al.}(1987)\citenamefont
  {Willett}, \citenamefont {Eisenstein}, \citenamefont {St\"ormer},
  \citenamefont {Tsui}, \citenamefont {Gossard},\ and\ \citenamefont
  {English}}]{Willett87}%
  \BibitemOpen
  \bibfield  {author} {\bibinfo {author} {\bibfnamefont {R.}~\bibnamefont
  {Willett}}, \bibinfo {author} {\bibfnamefont {J.~P.}\ \bibnamefont
  {Eisenstein}}, \bibinfo {author} {\bibfnamefont {H.~L.}\ \bibnamefont
  {St\"ormer}}, \bibinfo {author} {\bibfnamefont {D.~C.}\ \bibnamefont {Tsui}},
  \bibinfo {author} {\bibfnamefont {A.~C.}\ \bibnamefont {Gossard}},\ and\
  \bibinfo {author} {\bibfnamefont {J.~H.}\ \bibnamefont {English}},\
  }\bibfield  {title} {\bibinfo {title} {Observation of an even-denominator
  quantum number in the fractional quantum {Hall} effect},\ }\href
  {https://doi.org/10.1103/PhysRevLett.59.1776} {\bibfield  {journal} {\bibinfo
   {journal} {Phys. Rev. Lett.}\ }\textbf {\bibinfo {volume} {59}},\ \bibinfo
  {pages} {1776} (\bibinfo {year} {1987})}\BibitemShut {NoStop}%
\bibitem [{\citenamefont {Moore}\ and\ \citenamefont {Read}(1991)}]{Moore91}%
  \BibitemOpen
  \bibfield  {author} {\bibinfo {author} {\bibfnamefont {G.}~\bibnamefont
  {Moore}}\ and\ \bibinfo {author} {\bibfnamefont {N.}~\bibnamefont {Read}},\
  }\bibfield  {title} {\bibinfo {title} {Nonabelions in the fractional quantum
  {Hall} effect},\ }\href {https://doi.org/10.1016/0550-3213(91)90407-O}
  {\bibfield  {journal} {\bibinfo  {journal} {Nucl. Phys. B}\ }\textbf
  {\bibinfo {volume} {360}},\ \bibinfo {pages} {362 } (\bibinfo {year}
  {1991})}\BibitemShut {NoStop}%
\bibitem [{\citenamefont {Read}\ and\ \citenamefont {Green}(2000)}]{Read00}%
  \BibitemOpen
  \bibfield  {author} {\bibinfo {author} {\bibfnamefont {N.}~\bibnamefont
  {Read}}\ and\ \bibinfo {author} {\bibfnamefont {D.}~\bibnamefont {Green}},\
  }\bibfield  {title} {\bibinfo {title} {Paired states of fermions in two
  dimensions with breaking of parity and time-reversal symmetries and the
  fractional quantum {Hall} effect},\ }\href
  {https://doi.org/10.1103/PhysRevB.61.10267} {\bibfield  {journal} {\bibinfo
  {journal} {Phys. Rev. B}\ }\textbf {\bibinfo {volume} {61}},\ \bibinfo
  {pages} {10267} (\bibinfo {year} {2000})}\BibitemShut {NoStop}%
\bibitem [{\citenamefont {Levin}\ \emph {et~al.}(2007)\citenamefont {Levin},
  \citenamefont {Halperin},\ and\ \citenamefont {Rosenow}}]{Levin07}%
  \BibitemOpen
  \bibfield  {author} {\bibinfo {author} {\bibfnamefont {M.}~\bibnamefont
  {Levin}}, \bibinfo {author} {\bibfnamefont {B.~I.}\ \bibnamefont
  {Halperin}},\ and\ \bibinfo {author} {\bibfnamefont {B.}~\bibnamefont
  {Rosenow}},\ }\bibfield  {title} {\bibinfo {title} {Particle-hole symmetry
  and the {Pfaffian} state},\ }\href
  {https://doi.org/10.1103/PhysRevLett.99.236806} {\bibfield  {journal}
  {\bibinfo  {journal} {Phys. Rev. Lett.}\ }\textbf {\bibinfo {volume} {99}},\
  \bibinfo {pages} {236806} (\bibinfo {year} {2007})}\BibitemShut {NoStop}%
\bibitem [{\citenamefont {Lee}\ \emph {et~al.}(2007)\citenamefont {Lee},
  \citenamefont {Ryu}, \citenamefont {Nayak},\ and\ \citenamefont
  {Fisher}}]{Lee07}%
  \BibitemOpen
  \bibfield  {author} {\bibinfo {author} {\bibfnamefont {S.-S.}\ \bibnamefont
  {Lee}}, \bibinfo {author} {\bibfnamefont {S.}~\bibnamefont {Ryu}}, \bibinfo
  {author} {\bibfnamefont {C.}~\bibnamefont {Nayak}},\ and\ \bibinfo {author}
  {\bibfnamefont {M.~P.~A.}\ \bibnamefont {Fisher}},\ }\bibfield  {title}
  {\bibinfo {title} {Particle-hole symmetry and the $\nu=5/2$ quantum {Hall}
  state},\ }\href {https://doi.org/10.1103/PhysRevLett.99.236807} {\bibfield
  {journal} {\bibinfo  {journal} {Phys. Rev. Lett.}\ }\textbf {\bibinfo
  {volume} {99}},\ \bibinfo {pages} {236807} (\bibinfo {year}
  {2007})}\BibitemShut {NoStop}%
\bibitem [{\citenamefont {Balram}\ \emph {et~al.}(2018)\citenamefont {Balram},
  \citenamefont {Barkeshli},\ and\ \citenamefont {Rudner}}]{Balram18}%
  \BibitemOpen
  \bibfield  {author} {\bibinfo {author} {\bibfnamefont {A.~C.}\ \bibnamefont
  {Balram}}, \bibinfo {author} {\bibfnamefont {M.}~\bibnamefont {Barkeshli}},\
  and\ \bibinfo {author} {\bibfnamefont {M.~S.}\ \bibnamefont {Rudner}},\
  }\bibfield  {title} {\bibinfo {title} {Parton construction of a wave function
  in the anti-{Pfaffian} phase},\ }\href
  {https://doi.org/10.1103/PhysRevB.98.035127} {\bibfield  {journal} {\bibinfo
  {journal} {Phys. Rev. B}\ }\textbf {\bibinfo {volume} {98}},\ \bibinfo
  {pages} {035127} (\bibinfo {year} {2018})}\BibitemShut {NoStop}%
\bibitem [{\citenamefont {Sharma}\ \emph {et~al.}(2021)\citenamefont {Sharma},
  \citenamefont {Pu},\ and\ \citenamefont {Jain}}]{Sharma21}%
  \BibitemOpen
  \bibfield  {author} {\bibinfo {author} {\bibfnamefont {A.}~\bibnamefont
  {Sharma}}, \bibinfo {author} {\bibfnamefont {S.}~\bibnamefont {Pu}},\ and\
  \bibinfo {author} {\bibfnamefont {J.~K.}\ \bibnamefont {Jain}},\ }\bibfield
  {title} {\bibinfo {title} {Bardeen-cooper-schrieffer pairing of composite
  fermions},\ }\href {https://doi.org/10.1103/PhysRevB.104.205303} {\bibfield
  {journal} {\bibinfo  {journal} {Phys. Rev. B}\ }\textbf {\bibinfo {volume}
  {104}},\ \bibinfo {pages} {205303} (\bibinfo {year} {2021})}\BibitemShut
  {NoStop}%
\bibitem [{\citenamefont {Stern}\ \emph {et~al.}(2004)\citenamefont {Stern},
  \citenamefont {von Oppen},\ and\ \citenamefont {Mariani}}]{Stern04}%
  \BibitemOpen
  \bibfield  {author} {\bibinfo {author} {\bibfnamefont {A.}~\bibnamefont
  {Stern}}, \bibinfo {author} {\bibfnamefont {F.}~\bibnamefont {von Oppen}},\
  and\ \bibinfo {author} {\bibfnamefont {E.}~\bibnamefont {Mariani}},\
  }\bibfield  {title} {\bibinfo {title} {Geometric phases and quantum
  entanglement as building blocks for non-{Abelian} quasiparticle statistics},\
  }\href {https://doi.org/10.1103/PhysRevB.70.205338} {\bibfield  {journal}
  {\bibinfo  {journal} {Phys. Rev. B}\ }\textbf {\bibinfo {volume} {70}},\
  \bibinfo {pages} {205338} (\bibinfo {year} {2004})},\ \Eprint
  {https://arxiv.org/abs/cond-mat/0310273} {cond-mat/0310273} \BibitemShut
  {NoStop}%
\bibitem [{\citenamefont {Read}(2009)}]{Read09}%
  \BibitemOpen
  \bibfield  {author} {\bibinfo {author} {\bibfnamefont {N.}~\bibnamefont
  {Read}},\ }\bibfield  {title} {\bibinfo {title} {Non-abelian adiabatic
  statistics and {Hall} viscosity in quantum {Hall} states and
  ${p}_{x}+i{p}_{y}$ paired superfluids},\ }\href
  {https://doi.org/10.1103/PhysRevB.79.045308} {\bibfield  {journal} {\bibinfo
  {journal} {Phys. Rev. B}\ }\textbf {\bibinfo {volume} {79}},\ \bibinfo
  {pages} {045308} (\bibinfo {year} {2009})}\BibitemShut {NoStop}%
\bibitem [{\citenamefont {Bonderson}\ \emph {et~al.}(2011)\citenamefont
  {Bonderson}, \citenamefont {Gurarie},\ and\ \citenamefont
  {Nayak}}]{Bonderson11}%
  \BibitemOpen
  \bibfield  {author} {\bibinfo {author} {\bibfnamefont {P.}~\bibnamefont
  {Bonderson}}, \bibinfo {author} {\bibfnamefont {V.}~\bibnamefont {Gurarie}},\
  and\ \bibinfo {author} {\bibfnamefont {C.}~\bibnamefont {Nayak}},\ }\bibfield
   {title} {\bibinfo {title} {Plasma analogy and non-{Abelian} statistics for
  ising-type quantum {Hall} states},\ }\href
  {https://doi.org/10.1103/PhysRevB.83.075303} {\bibfield  {journal} {\bibinfo
  {journal} {Phys. Rev. B}\ }\textbf {\bibinfo {volume} {83}},\ \bibinfo
  {pages} {075303} (\bibinfo {year} {2011})}\BibitemShut {NoStop}%
\bibitem [{\citenamefont {Das~Sarma}\ \emph {et~al.}(2005)\citenamefont
  {Das~Sarma}, \citenamefont {Freedman},\ and\ \citenamefont
  {Nayak}}]{DasSarma05}%
  \BibitemOpen
  \bibfield  {author} {\bibinfo {author} {\bibfnamefont {S.}~\bibnamefont
  {Das~Sarma}}, \bibinfo {author} {\bibfnamefont {M.}~\bibnamefont
  {Freedman}},\ and\ \bibinfo {author} {\bibfnamefont {C.}~\bibnamefont
  {Nayak}},\ }\bibfield  {title} {\bibinfo {title} {Topologically protected
  qubits from a possible non-{Abelian} fractional quantum {Hall} state},\
  }\href {https://doi.org/10.1103/PhysRevLett.94.166802} {\bibfield  {journal}
  {\bibinfo  {journal} {Phys. Rev. Lett.}\ }\textbf {\bibinfo {volume} {94}},\
  \bibinfo {pages} {166802} (\bibinfo {year} {2005})}\BibitemShut {NoStop}%
\bibitem [{\citenamefont {Das~Sarma}\ \emph {et~al.}(2006)\citenamefont
  {Das~Sarma}, \citenamefont {Nayak},\ and\ \citenamefont
  {Tewari}}]{DasSarma06}%
  \BibitemOpen
  \bibfield  {author} {\bibinfo {author} {\bibfnamefont {S.}~\bibnamefont
  {Das~Sarma}}, \bibinfo {author} {\bibfnamefont {C.}~\bibnamefont {Nayak}},\
  and\ \bibinfo {author} {\bibfnamefont {S.}~\bibnamefont {Tewari}},\
  }\bibfield  {title} {\bibinfo {title} {Proposal to stabilize and detect
  half-quantum vortices in strontium ruthenate thin films: Non-{Abelian}
  braiding statistics of vortices in a ${p}_{x}+i{p}_{y}$ superconductor},\
  }\href {https://doi.org/10.1103/PhysRevB.73.220502} {\bibfield  {journal}
  {\bibinfo  {journal} {Phys. Rev. B}\ }\textbf {\bibinfo {volume} {73}},\
  \bibinfo {pages} {220502} (\bibinfo {year} {2006})}\BibitemShut {NoStop}%
\bibitem [{\citenamefont {Stern}\ and\ \citenamefont
  {Halperin}(2006)}]{Stern06}%
  \BibitemOpen
  \bibfield  {author} {\bibinfo {author} {\bibfnamefont {A.}~\bibnamefont
  {Stern}}\ and\ \bibinfo {author} {\bibfnamefont {B.~I.}\ \bibnamefont
  {Halperin}},\ }\bibfield  {title} {\bibinfo {title} {Proposed experiments to
  probe the non-abelian $\nu=5/2$ quantum {Hall} state},\ }\href
  {https://doi.org/10.1103/PhysRevLett.96.016802} {\bibfield  {journal}
  {\bibinfo  {journal} {Phys. Rev. Lett.}\ }\textbf {\bibinfo {volume} {96}},\
  \bibinfo {pages} {016802} (\bibinfo {year} {2006})}\BibitemShut {NoStop}%
\bibitem [{\citenamefont {Nayak}\ \emph {et~al.}(2008)\citenamefont {Nayak},
  \citenamefont {Simon}, \citenamefont {Stern}, \citenamefont {Freedman},\ and\
  \citenamefont {Das~Sarma}}]{Nayak08}%
  \BibitemOpen
  \bibfield  {author} {\bibinfo {author} {\bibfnamefont {C.}~\bibnamefont
  {Nayak}}, \bibinfo {author} {\bibfnamefont {S.~H.}\ \bibnamefont {Simon}},
  \bibinfo {author} {\bibfnamefont {A.}~\bibnamefont {Stern}}, \bibinfo
  {author} {\bibfnamefont {M.}~\bibnamefont {Freedman}},\ and\ \bibinfo
  {author} {\bibfnamefont {S.}~\bibnamefont {Das~Sarma}},\ }\bibfield  {title}
  {\bibinfo {title} {Non-{Abelian} anyons and topological quantum
  computation},\ }\href {https://doi.org/10.1103/RevModPhys.80.1083} {\bibfield
   {journal} {\bibinfo  {journal} {Rev. Mod. Phys.}\ }\textbf {\bibinfo
  {volume} {80}},\ \bibinfo {pages} {1083} (\bibinfo {year}
  {2008})}\BibitemShut {NoStop}%
\bibitem [{\citenamefont {{Stern}}(2010)}]{Stern10}%
  \BibitemOpen
  \bibfield  {author} {\bibinfo {author} {\bibfnamefont {A.}~\bibnamefont
  {{Stern}}},\ }\bibfield  {title} {\bibinfo {title} {{Non-{Abelian} states of
  matter}},\ }\href {https://doi.org/10.1038/nature08915} {\bibfield  {journal}
  {\bibinfo  {journal} {Nature}\ }\textbf {\bibinfo {volume} {464}},\ \bibinfo
  {pages} {187} (\bibinfo {year} {2010})}\BibitemShut {NoStop}%
\bibitem [{\citenamefont {Sarma}\ \emph {et~al.}(2015)\citenamefont {Sarma},
  \citenamefont {Freedman},\ and\ \citenamefont {Nayak}}]{DasSarma15}%
  \BibitemOpen
  \bibfield  {author} {\bibinfo {author} {\bibfnamefont {S.~D.}\ \bibnamefont
  {Sarma}}, \bibinfo {author} {\bibfnamefont {M.}~\bibnamefont {Freedman}},\
  and\ \bibinfo {author} {\bibfnamefont {C.}~\bibnamefont {Nayak}},\ }\bibfield
   {title} {\bibinfo {title} {Majorana zero modes and topological quantum
  computation},\ }\href {https://www.nature.com/articles/npjqi20151} {\bibfield
   {journal} {\bibinfo  {journal} {npj Quantum Information}\ }\textbf {\bibinfo
  {volume} {1}},\ \bibinfo {pages} {1} (\bibinfo {year} {2015})}\BibitemShut
  {NoStop}%
\bibitem [{\citenamefont {Suen}\ \emph {et~al.}(1992)\citenamefont {Suen},
  \citenamefont {Engel}, \citenamefont {Santos}, \citenamefont {Shayegan},\
  and\ \citenamefont {Tsui}}]{Suen92}%
  \BibitemOpen
  \bibfield  {author} {\bibinfo {author} {\bibfnamefont {Y.~W.}\ \bibnamefont
  {Suen}}, \bibinfo {author} {\bibfnamefont {L.~W.}\ \bibnamefont {Engel}},
  \bibinfo {author} {\bibfnamefont {M.~B.}\ \bibnamefont {Santos}}, \bibinfo
  {author} {\bibfnamefont {M.}~\bibnamefont {Shayegan}},\ and\ \bibinfo
  {author} {\bibfnamefont {D.~C.}\ \bibnamefont {Tsui}},\ }\bibfield  {title}
  {\bibinfo {title} {Observation of a $\nu=1/2$ fractional quantum {Hall} state
  in a double-layer electron system},\ }\href
  {https://doi.org/10.1103/PhysRevLett.68.1379} {\bibfield  {journal} {\bibinfo
   {journal} {Phys. Rev. Lett.}\ }\textbf {\bibinfo {volume} {68}},\ \bibinfo
  {pages} {1379} (\bibinfo {year} {1992})}\BibitemShut {NoStop}%
\bibitem [{\citenamefont {Shabani}\ \emph
  {et~al.}(2009{\natexlab{a}})\citenamefont {Shabani}, \citenamefont {Gokmen},\
  and\ \citenamefont {Shayegan}}]{Shabani09}%
  \BibitemOpen
  \bibfield  {author} {\bibinfo {author} {\bibfnamefont {J.}~\bibnamefont
  {Shabani}}, \bibinfo {author} {\bibfnamefont {T.}~\bibnamefont {Gokmen}},\
  and\ \bibinfo {author} {\bibfnamefont {M.}~\bibnamefont {Shayegan}},\
  }\bibfield  {title} {\bibinfo {title} {Correlated states of electrons in wide
  quantum wells at low fillings: The role of charge distribution symmetry},\
  }\href {https://doi.org/10.1103/PhysRevLett.103.046805} {\bibfield  {journal}
  {\bibinfo  {journal} {Phys. Rev. Lett.}\ }\textbf {\bibinfo {volume} {103}},\
  \bibinfo {pages} {046805} (\bibinfo {year} {2009}{\natexlab{a}})}\BibitemShut
  {NoStop}%
\bibitem [{\citenamefont {Sharma}\ \emph {et~al.}(2024)\citenamefont {Sharma},
  \citenamefont {Balram},\ and\ \citenamefont {Jain}}]{Sharma24}%
  \BibitemOpen
  \bibfield  {author} {\bibinfo {author} {\bibfnamefont {A.}~\bibnamefont
  {Sharma}}, \bibinfo {author} {\bibfnamefont {A.~C.}\ \bibnamefont {Balram}},\
  and\ \bibinfo {author} {\bibfnamefont {J.~K.}\ \bibnamefont {Jain}},\
  }\bibfield  {title} {\bibinfo {title} {Composite-fermion pairing at
  half-filled and quarter-filled lowest landau level},\ }\href
  {https://doi.org/10.1103/PhysRevB.109.035306} {\bibfield  {journal} {\bibinfo
   {journal} {Phys. Rev. B}\ }\textbf {\bibinfo {volume} {109}},\ \bibinfo
  {pages} {035306} (\bibinfo {year} {2024})}\BibitemShut {NoStop}%
\bibitem [{\citenamefont {Shabani}\ \emph {et~al.}(2013)\citenamefont
  {Shabani}, \citenamefont {Liu}, \citenamefont {Shayegan}, \citenamefont
  {Pfeiffer}, \citenamefont {West},\ and\ \citenamefont {Baldwin}}]{Shabani13}%
  \BibitemOpen
  \bibfield  {author} {\bibinfo {author} {\bibfnamefont {J.}~\bibnamefont
  {Shabani}}, \bibinfo {author} {\bibfnamefont {Y.}~\bibnamefont {Liu}},
  \bibinfo {author} {\bibfnamefont {M.}~\bibnamefont {Shayegan}}, \bibinfo
  {author} {\bibfnamefont {L.~N.}\ \bibnamefont {Pfeiffer}}, \bibinfo {author}
  {\bibfnamefont {K.~W.}\ \bibnamefont {West}},\ and\ \bibinfo {author}
  {\bibfnamefont {K.~W.}\ \bibnamefont {Baldwin}},\ }\bibfield  {title}
  {\bibinfo {title} {Phase diagrams for the stability of the
  $\ensuremath{\nu}=\frac{1}{2}$ fractional quantum {Hall} effect in electron
  systems confined to symmetric, wide {Ga}{As} quantum wells},\ }\href
  {https://doi.org/10.1103/PhysRevB.88.245413} {\bibfield  {journal} {\bibinfo
  {journal} {Phys. Rev. B}\ }\textbf {\bibinfo {volume} {88}},\ \bibinfo
  {pages} {245413} (\bibinfo {year} {2013})}\BibitemShut {NoStop}%
\bibitem [{\citenamefont {Wang}\ \emph {et~al.}(2022)\citenamefont {Wang},
  \citenamefont {Gupta}, \citenamefont {Singh}, \citenamefont {Chung},
  \citenamefont {Pfeiffer}, \citenamefont {West}, \citenamefont {Baldwin},
  \citenamefont {Winkler},\ and\ \citenamefont {Shayegan}}]{Wang22A}%
  \BibitemOpen
  \bibfield  {author} {\bibinfo {author} {\bibfnamefont {C.}~\bibnamefont
  {Wang}}, \bibinfo {author} {\bibfnamefont {A.}~\bibnamefont {Gupta}},
  \bibinfo {author} {\bibfnamefont {S.~K.}\ \bibnamefont {Singh}}, \bibinfo
  {author} {\bibfnamefont {Y.~J.}\ \bibnamefont {Chung}}, \bibinfo {author}
  {\bibfnamefont {L.~N.}\ \bibnamefont {Pfeiffer}}, \bibinfo {author}
  {\bibfnamefont {K.~W.}\ \bibnamefont {West}}, \bibinfo {author}
  {\bibfnamefont {K.~W.}\ \bibnamefont {Baldwin}}, \bibinfo {author}
  {\bibfnamefont {R.}~\bibnamefont {Winkler}},\ and\ \bibinfo {author}
  {\bibfnamefont {M.}~\bibnamefont {Shayegan}},\ }\bibfield  {title} {\bibinfo
  {title} {Even-denominator fractional quantum hall state at filling factor
  $\ensuremath{\nu}=3/4$},\ }\href
  {https://doi.org/10.1103/PhysRevLett.129.156801} {\bibfield  {journal}
  {\bibinfo  {journal} {Phys. Rev. Lett.}\ }\textbf {\bibinfo {volume} {129}},\
  \bibinfo {pages} {156801} (\bibinfo {year} {2022})}\BibitemShut {NoStop}%
\bibitem [{\citenamefont {Zhao}\ \emph {et~al.}(2023)\citenamefont {Zhao},
  \citenamefont {Balram},\ and\ \citenamefont {Jain}}]{Zhao23}%
  \BibitemOpen
  \bibfield  {author} {\bibinfo {author} {\bibfnamefont {T.}~\bibnamefont
  {Zhao}}, \bibinfo {author} {\bibfnamefont {A.~C.}\ \bibnamefont {Balram}},\
  and\ \bibinfo {author} {\bibfnamefont {J.~K.}\ \bibnamefont {Jain}},\
  }\bibfield  {title} {\bibinfo {title} {Composite fermion pairing induced by
  landau level mixing},\ }\href
  {https://doi.org/10.1103/PhysRevLett.130.186302} {\bibfield  {journal}
  {\bibinfo  {journal} {Phys. Rev. Lett.}\ }\textbf {\bibinfo {volume} {130}},\
  \bibinfo {pages} {186302} (\bibinfo {year} {2023})}\BibitemShut {NoStop}%
\bibitem [{\citenamefont {Wang}\ \emph {et~al.}(2023)\citenamefont {Wang},
  \citenamefont {Gupta}, \citenamefont {Singh}, \citenamefont {Madathil},
  \citenamefont {Chung}, \citenamefont {Pfeiffer}, \citenamefont {Baldwin},
  \citenamefont {Winkler},\ and\ \citenamefont {Shayegan}}]{Wang23}%
  \BibitemOpen
  \bibfield  {author} {\bibinfo {author} {\bibfnamefont {C.}~\bibnamefont
  {Wang}}, \bibinfo {author} {\bibfnamefont {A.}~\bibnamefont {Gupta}},
  \bibinfo {author} {\bibfnamefont {S.~K.}\ \bibnamefont {Singh}}, \bibinfo
  {author} {\bibfnamefont {P.~T.}\ \bibnamefont {Madathil}}, \bibinfo {author}
  {\bibfnamefont {Y.~J.}\ \bibnamefont {Chung}}, \bibinfo {author}
  {\bibfnamefont {L.~N.}\ \bibnamefont {Pfeiffer}}, \bibinfo {author}
  {\bibfnamefont {K.~W.}\ \bibnamefont {Baldwin}}, \bibinfo {author}
  {\bibfnamefont {R.}~\bibnamefont {Winkler}},\ and\ \bibinfo {author}
  {\bibfnamefont {M.}~\bibnamefont {Shayegan}},\ }\bibfield  {title} {\bibinfo
  {title} {Fractional quantum hall state at filling factor
  $\ensuremath{\nu}=1/4$ in ultra-high-quality gaas two-dimensional hole
  systems},\ }\href {https://doi.org/10.1103/PhysRevLett.131.266502} {\bibfield
   {journal} {\bibinfo  {journal} {Phys. Rev. Lett.}\ }\textbf {\bibinfo
  {volume} {131}},\ \bibinfo {pages} {266502} (\bibinfo {year}
  {2023})}\BibitemShut {NoStop}%
\bibitem [{\citenamefont {Kim}\ \emph {et~al.}(2019)\citenamefont {Kim},
  \citenamefont {Balram}, \citenamefont {Taniguchi}, \citenamefont {Watanabe},
  \citenamefont {Jain},\ and\ \citenamefont {Smet}}]{Kim19}%
  \BibitemOpen
  \bibfield  {author} {\bibinfo {author} {\bibfnamefont {Y.}~\bibnamefont
  {Kim}}, \bibinfo {author} {\bibfnamefont {A.~C.}\ \bibnamefont {Balram}},
  \bibinfo {author} {\bibfnamefont {T.}~\bibnamefont {Taniguchi}}, \bibinfo
  {author} {\bibfnamefont {K.}~\bibnamefont {Watanabe}}, \bibinfo {author}
  {\bibfnamefont {J.~K.}\ \bibnamefont {Jain}},\ and\ \bibinfo {author}
  {\bibfnamefont {J.~H.}\ \bibnamefont {Smet}},\ }\bibfield  {title} {\bibinfo
  {title} {Even denominator fractional quantum {Hall} states in higher {Landau}
  levels of graphene},\ }\href {https://doi.org/10.1038/s41567-018-0355-x}
  {\bibfield  {journal} {\bibinfo  {journal} {Nature Physics}\ }\textbf
  {\bibinfo {volume} {15}},\ \bibinfo {pages} {154} (\bibinfo {year}
  {2019})}\BibitemShut {NoStop}%
\bibitem [{\citenamefont {Sharma}\ \emph {et~al.}(2023)\citenamefont {Sharma},
  \citenamefont {Pu}, \citenamefont {Balram},\ and\ \citenamefont
  {Jain}}]{Sharma23}%
  \BibitemOpen
  \bibfield  {author} {\bibinfo {author} {\bibfnamefont {A.}~\bibnamefont
  {Sharma}}, \bibinfo {author} {\bibfnamefont {S.}~\bibnamefont {Pu}}, \bibinfo
  {author} {\bibfnamefont {A.~C.}\ \bibnamefont {Balram}},\ and\ \bibinfo
  {author} {\bibfnamefont {J.~K.}\ \bibnamefont {Jain}},\ }\bibfield  {title}
  {\bibinfo {title} {Fractional quantum {Hall} effect with unconventional
  pairing in monolayer graphene},\ }\href
  {https://doi.org/10.1103/PhysRevLett.130.126201} {\bibfield  {journal}
  {\bibinfo  {journal} {Phys. Rev. Lett.}\ }\textbf {\bibinfo {volume} {130}},\
  \bibinfo {pages} {126201} (\bibinfo {year} {2023})}\BibitemShut {NoStop}%
\bibitem [{\citenamefont {Domaretskiy}\ \emph {et~al.}(2026)\citenamefont
  {Domaretskiy}, \citenamefont {Hayward}, \citenamefont {Nguyen}, \citenamefont
  {Benaglia}, \citenamefont {Indykiewicz}, \citenamefont {Vignaud},
  \citenamefont {Zhang}, \citenamefont {Watanabe}, \citenamefont {Taniguchi},
  \citenamefont {Fal'ko} \emph {et~al.}}]{Domaretskiy26}%
  \BibitemOpen
  \bibfield  {author} {\bibinfo {author} {\bibfnamefont {D.}~\bibnamefont
  {Domaretskiy}}, \bibinfo {author} {\bibfnamefont {N.}~\bibnamefont
  {Hayward}}, \bibinfo {author} {\bibfnamefont {V.~H.}\ \bibnamefont {Nguyen}},
  \bibinfo {author} {\bibfnamefont {S.}~\bibnamefont {Benaglia}}, \bibinfo
  {author} {\bibfnamefont {K.}~\bibnamefont {Indykiewicz}}, \bibinfo {author}
  {\bibfnamefont {H.}~\bibnamefont {Vignaud}}, \bibinfo {author} {\bibfnamefont
  {J.}~\bibnamefont {Zhang}}, \bibinfo {author} {\bibfnamefont
  {K.}~\bibnamefont {Watanabe}}, \bibinfo {author} {\bibfnamefont
  {T.}~\bibnamefont {Taniguchi}}, \bibinfo {author} {\bibfnamefont
  {V.}~\bibnamefont {Fal'ko}}, \emph {et~al.},\ }\bibfield  {title} {\bibinfo
  {title} {Deep-uv bleaching of charge disorder in encapsulated graphene},\
  }\href@noop {} {\bibfield  {journal} {\bibinfo  {journal} {arXiv preprint
  arXiv:2603.29891}\ } (\bibinfo {year} {2026})}\BibitemShut {NoStop}%
\bibitem [{\citenamefont {Ki}\ \emph {et~al.}(2014)\citenamefont {Ki},
  \citenamefont {Fal'ko}, \citenamefont {Abanin},\ and\ \citenamefont
  {Morpurgo}}]{Ki14}%
  \BibitemOpen
  \bibfield  {author} {\bibinfo {author} {\bibfnamefont {D.-K.}\ \bibnamefont
  {Ki}}, \bibinfo {author} {\bibfnamefont {V.~I.}\ \bibnamefont {Fal'ko}},
  \bibinfo {author} {\bibfnamefont {D.~A.}\ \bibnamefont {Abanin}},\ and\
  \bibinfo {author} {\bibfnamefont {A.~F.}\ \bibnamefont {Morpurgo}},\
  }\bibfield  {title} {\bibinfo {title} {Observation of even denominator
  fractional quantum {Hall} effect in suspended bilayer graphene},\ }\href
  {https://doi.org/10.1021/nl5003922} {\bibfield  {journal} {\bibinfo
  {journal} {Nano Letters}\ }\textbf {\bibinfo {volume} {14}},\ \bibinfo
  {pages} {2135} (\bibinfo {year} {2014})},\ \bibinfo {note} {pMID: 24611523},\
  \Eprint {https://arxiv.org/abs/http://dx.doi.org/10.1021/nl5003922}
  {http://dx.doi.org/10.1021/nl5003922} \BibitemShut {NoStop}%
\bibitem [{\citenamefont {Zibrov}\ \emph {et~al.}(2017)\citenamefont {Zibrov},
  \citenamefont {Kometter}, \citenamefont {Zhou}, \citenamefont {Spanton},
  \citenamefont {Taniguchi}, \citenamefont {Watanabe}, \citenamefont
  {Zaletel},\ and\ \citenamefont {Young}}]{Zibrov17}%
  \BibitemOpen
  \bibfield  {author} {\bibinfo {author} {\bibfnamefont {A.~A.}\ \bibnamefont
  {Zibrov}}, \bibinfo {author} {\bibfnamefont {C.}~\bibnamefont {Kometter}},
  \bibinfo {author} {\bibfnamefont {H.}~\bibnamefont {Zhou}}, \bibinfo {author}
  {\bibfnamefont {E.}~\bibnamefont {Spanton}}, \bibinfo {author} {\bibfnamefont
  {T.}~\bibnamefont {Taniguchi}}, \bibinfo {author} {\bibfnamefont
  {K.}~\bibnamefont {Watanabe}}, \bibinfo {author} {\bibfnamefont
  {M.}~\bibnamefont {Zaletel}},\ and\ \bibinfo {author} {\bibfnamefont
  {A.}~\bibnamefont {Young}},\ }\bibfield  {title} {\bibinfo {title} {Tunable
  interacting composite fermion phases in a half-filled bilayer-graphene landau
  level},\ }\href@noop {} {\bibfield  {journal} {\bibinfo  {journal} {Nature}\
  }\textbf {\bibinfo {volume} {549}},\ \bibinfo {pages} {360} (\bibinfo {year}
  {2017})}\BibitemShut {NoStop}%
\bibitem [{\citenamefont {Li}\ \emph {et~al.}(2017)\citenamefont {Li},
  \citenamefont {Tan}, \citenamefont {Chen}, \citenamefont {Zeng},
  \citenamefont {Taniguchi}, \citenamefont {Watanabe}, \citenamefont {Hone},\
  and\ \citenamefont {Dean}}]{Li17}%
  \BibitemOpen
  \bibfield  {author} {\bibinfo {author} {\bibfnamefont {J.~I.~A.}\
  \bibnamefont {Li}}, \bibinfo {author} {\bibfnamefont {C.}~\bibnamefont
  {Tan}}, \bibinfo {author} {\bibfnamefont {S.}~\bibnamefont {Chen}}, \bibinfo
  {author} {\bibfnamefont {Y.}~\bibnamefont {Zeng}}, \bibinfo {author}
  {\bibfnamefont {T.}~\bibnamefont {Taniguchi}}, \bibinfo {author}
  {\bibfnamefont {K.}~\bibnamefont {Watanabe}}, \bibinfo {author}
  {\bibfnamefont {J.}~\bibnamefont {Hone}},\ and\ \bibinfo {author}
  {\bibfnamefont {C.~R.}\ \bibnamefont {Dean}},\ }\bibfield  {title} {\bibinfo
  {title} {Even denominator fractional quantum {Hall} states in bilayer
  graphene},\ }\href {https://doi.org/10.1126/science.aao2521} {\bibfield
  {journal} {\bibinfo  {journal} {Science}\ }\textbf {\bibinfo {volume}
  {358}},\ \bibinfo {pages} {648} (\bibinfo {year} {2017})}\BibitemShut
  {NoStop}%
\bibitem [{\citenamefont {Li}\ \emph {et~al.}(2018)\citenamefont {Li},
  \citenamefont {Tupikov}, \citenamefont {Watanabe}, \citenamefont
  {Taniguchi},\ and\ \citenamefont {Zhu}}]{Li18}%
  \BibitemOpen
  \bibfield  {author} {\bibinfo {author} {\bibfnamefont {J.}~\bibnamefont
  {Li}}, \bibinfo {author} {\bibfnamefont {Y.}~\bibnamefont {Tupikov}},
  \bibinfo {author} {\bibfnamefont {K.}~\bibnamefont {Watanabe}}, \bibinfo
  {author} {\bibfnamefont {T.}~\bibnamefont {Taniguchi}},\ and\ \bibinfo
  {author} {\bibfnamefont {J.}~\bibnamefont {Zhu}},\ }\bibfield  {title}
  {\bibinfo {title} {Effective landau level diagram of bilayer graphene},\
  }\href {https://doi.org/10.1103/PhysRevLett.120.047701} {\bibfield  {journal}
  {\bibinfo  {journal} {Phys. Rev. Lett.}\ }\textbf {\bibinfo {volume} {120}},\
  \bibinfo {pages} {047701} (\bibinfo {year} {2018})}\BibitemShut {NoStop}%
\bibitem [{\citenamefont {Wang}\ \emph {et~al.}(2025)\citenamefont {Wang},
  \citenamefont {Madathil}, \citenamefont {Singh}, \citenamefont {Gupta},
  \citenamefont {Chung}, \citenamefont {Pfeiffer}, \citenamefont {Baldwin},\
  and\ \citenamefont {Shayegan}}]{Wang25}%
  \BibitemOpen
  \bibfield  {author} {\bibinfo {author} {\bibfnamefont {C.}~\bibnamefont
  {Wang}}, \bibinfo {author} {\bibfnamefont {P.~T.}\ \bibnamefont {Madathil}},
  \bibinfo {author} {\bibfnamefont {S.~K.}\ \bibnamefont {Singh}}, \bibinfo
  {author} {\bibfnamefont {A.}~\bibnamefont {Gupta}}, \bibinfo {author}
  {\bibfnamefont {Y.~J.}\ \bibnamefont {Chung}}, \bibinfo {author}
  {\bibfnamefont {L.~N.}\ \bibnamefont {Pfeiffer}}, \bibinfo {author}
  {\bibfnamefont {K.~W.}\ \bibnamefont {Baldwin}},\ and\ \bibinfo {author}
  {\bibfnamefont {M.}~\bibnamefont {Shayegan}},\ }\bibfield  {title} {\bibinfo
  {title} {Developing fractional quantum hall states at even-denominator
  fillings $1/6$ and $1/8$},\ }\href
  {https://doi.org/10.1103/PhysRevLett.134.046502} {\bibfield  {journal}
  {\bibinfo  {journal} {Phys. Rev. Lett.}\ }\textbf {\bibinfo {volume} {134}},\
  \bibinfo {pages} {046502} (\bibinfo {year} {2025})}\BibitemShut {NoStop}%
\bibitem [{\citenamefont {Balram}\ \emph {et~al.}(2025)\citenamefont {Balram},
  \citenamefont {Sharma},\ and\ \citenamefont {Jain}}]{Balram25}%
  \BibitemOpen
  \bibfield  {author} {\bibinfo {author} {\bibfnamefont {A.~C.}\ \bibnamefont
  {Balram}}, \bibinfo {author} {\bibfnamefont {A.}~\bibnamefont {Sharma}},\
  and\ \bibinfo {author} {\bibfnamefont {J.~K.}\ \bibnamefont {Jain}},\
  }\bibfield  {title} {\bibinfo {title} {Interplay of superconducting,
  metallic, and crystalline states of composite fermions at
  $\ensuremath{\nu}=\frac{1}{6}$ in wide quantum wells},\ }\href
  {https://doi.org/10.1103/ntrz-d86j} {\bibfield  {journal} {\bibinfo
  {journal} {Phys. Rev. B}\ }\textbf {\bibinfo {volume} {112}},\ \bibinfo
  {pages} {035118} (\bibinfo {year} {2025})}\BibitemShut {NoStop}%
\bibitem [{\citenamefont {Shabani}\ \emph
  {et~al.}(2009{\natexlab{b}})\citenamefont {Shabani}, \citenamefont {Gokmen},
  \citenamefont {Chiu},\ and\ \citenamefont {Shayegan}}]{Shabani09A}%
  \BibitemOpen
  \bibfield  {author} {\bibinfo {author} {\bibfnamefont {J.}~\bibnamefont
  {Shabani}}, \bibinfo {author} {\bibfnamefont {T.}~\bibnamefont {Gokmen}},
  \bibinfo {author} {\bibfnamefont {Y.~T.}\ \bibnamefont {Chiu}},\ and\
  \bibinfo {author} {\bibfnamefont {M.}~\bibnamefont {Shayegan}},\ }\bibfield
  {title} {\bibinfo {title} {Evidence for developing fractional quantum {Hall}
  states at even denominator $1/2$ and $1/4$ fillings in asymmetric wide
  quantum wells},\ }\href {https://doi.org/10.1103/PhysRevLett.103.256802}
  {\bibfield  {journal} {\bibinfo  {journal} {Phys. Rev. Lett.}\ }\textbf
  {\bibinfo {volume} {103}},\ \bibinfo {pages} {256802} (\bibinfo {year}
  {2009}{\natexlab{b}})}\BibitemShut {NoStop}%
\bibitem [{\citenamefont {Faugno}\ \emph {et~al.}(2019)\citenamefont {Faugno},
  \citenamefont {Balram}, \citenamefont {Barkeshli},\ and\ \citenamefont
  {Jain}}]{Faugno19}%
  \BibitemOpen
  \bibfield  {author} {\bibinfo {author} {\bibfnamefont {W.~N.}\ \bibnamefont
  {Faugno}}, \bibinfo {author} {\bibfnamefont {A.~C.}\ \bibnamefont {Balram}},
  \bibinfo {author} {\bibfnamefont {M.}~\bibnamefont {Barkeshli}},\ and\
  \bibinfo {author} {\bibfnamefont {J.~K.}\ \bibnamefont {Jain}},\ }\bibfield
  {title} {\bibinfo {title} {Prediction of a non-{Abelian} fractional quantum
  {Hall} state with $f$-wave pairing of composite fermions in wide quantum
  wells},\ }\href {https://doi.org/10.1103/PhysRevLett.123.016802} {\bibfield
  {journal} {\bibinfo  {journal} {Phys. Rev. Lett.}\ }\textbf {\bibinfo
  {volume} {123}},\ \bibinfo {pages} {016802} (\bibinfo {year}
  {2019})}\BibitemShut {NoStop}%
\bibitem [{\citenamefont {Dolev}\ \emph {et~al.}(2008)\citenamefont {Dolev},
  \citenamefont {Heiblum}, \citenamefont {Stern}, \citenamefont {Umansky},\
  and\ \citenamefont {Mahalu}}]{Dolev08}%
  \BibitemOpen
  \bibfield  {author} {\bibinfo {author} {\bibfnamefont {M.}~\bibnamefont
  {Dolev}}, \bibinfo {author} {\bibfnamefont {M.}~\bibnamefont {Heiblum}},
  \bibinfo {author} {\bibfnamefont {A.}~\bibnamefont {Stern}}, \bibinfo
  {author} {\bibfnamefont {V.}~\bibnamefont {Umansky}},\ and\ \bibinfo {author}
  {\bibfnamefont {D.}~\bibnamefont {Mahalu}},\ }\bibfield  {title} {\bibinfo
  {title} {Observation of a quarter of an electron charge at the $\nu=5/2$
  quantum {Hall} state},\ }\href {https://doi.org/10.1038/nature06855}
  {\bibfield  {journal} {\bibinfo  {journal} {Nature}\ }\textbf {\bibinfo
  {volume} {452}},\ \bibinfo {pages} {829} (\bibinfo {year}
  {2008})}\BibitemShut {NoStop}%
\bibitem [{\citenamefont {Dolev}\ \emph {et~al.}(2010)\citenamefont {Dolev},
  \citenamefont {Gross}, \citenamefont {Chung}, \citenamefont {Heiblum},
  \citenamefont {Umansky},\ and\ \citenamefont {Mahalu}}]{Dolev10}%
  \BibitemOpen
  \bibfield  {author} {\bibinfo {author} {\bibfnamefont {M.}~\bibnamefont
  {Dolev}}, \bibinfo {author} {\bibfnamefont {Y.}~\bibnamefont {Gross}},
  \bibinfo {author} {\bibfnamefont {Y.~C.}\ \bibnamefont {Chung}}, \bibinfo
  {author} {\bibfnamefont {M.}~\bibnamefont {Heiblum}}, \bibinfo {author}
  {\bibfnamefont {V.}~\bibnamefont {Umansky}},\ and\ \bibinfo {author}
  {\bibfnamefont {D.}~\bibnamefont {Mahalu}},\ }\bibfield  {title} {\bibinfo
  {title} {Dependence of the tunneling quasiparticle charge determined via shot
  noise measurements on the tunneling barrier and energetics},\ }\href
  {https://doi.org/10.1103/PhysRevB.81.161303} {\bibfield  {journal} {\bibinfo
  {journal} {Phys. Rev. B}\ }\textbf {\bibinfo {volume} {81}},\ \bibinfo
  {pages} {161303} (\bibinfo {year} {2010})}\BibitemShut {NoStop}%
\bibitem [{\citenamefont {Kim}\ \emph {et~al.}(2026)\citenamefont {Kim},
  \citenamefont {Dev}, \citenamefont {Shaer}, \citenamefont {Kumar},
  \citenamefont {Ilin}, \citenamefont {Haug}, \citenamefont {Iskoz},
  \citenamefont {Watanabe}, \citenamefont {Taniguchi}, \citenamefont {Mross},
  \citenamefont {Stern},\ and\ \citenamefont {Ronen}}]{Kim26}%
  \BibitemOpen
  \bibfield  {author} {\bibinfo {author} {\bibfnamefont {J.}~\bibnamefont
  {Kim}}, \bibinfo {author} {\bibfnamefont {H.}~\bibnamefont {Dev}}, \bibinfo
  {author} {\bibfnamefont {A.}~\bibnamefont {Shaer}}, \bibinfo {author}
  {\bibfnamefont {R.}~\bibnamefont {Kumar}}, \bibinfo {author} {\bibfnamefont
  {A.}~\bibnamefont {Ilin}}, \bibinfo {author} {\bibfnamefont {A.}~\bibnamefont
  {Haug}}, \bibinfo {author} {\bibfnamefont {S.}~\bibnamefont {Iskoz}},
  \bibinfo {author} {\bibfnamefont {K.}~\bibnamefont {Watanabe}}, \bibinfo
  {author} {\bibfnamefont {T.}~\bibnamefont {Taniguchi}}, \bibinfo {author}
  {\bibfnamefont {D.~F.}\ \bibnamefont {Mross}}, \bibinfo {author}
  {\bibfnamefont {A.}~\bibnamefont {Stern}},\ and\ \bibinfo {author}
  {\bibfnamefont {Y.}~\bibnamefont {Ronen}},\ }\bibfield  {title} {\bibinfo
  {title} {Aharonov-{Bohm} interference in even-denominator fractional quantum
  {Hall} states},\ }\href {https://doi.org/10.1038/s41586-025-09891-2}
  {\bibfield  {journal} {\bibinfo  {journal} {Nature}\ }\textbf {\bibinfo
  {volume} {649}},\ \bibinfo {pages} {323} (\bibinfo {year}
  {2026})}\BibitemShut {NoStop}%
\bibitem [{\citenamefont {Alkalay}\ \emph {et~al.}(2026)\citenamefont
  {Alkalay}, \citenamefont {Hajigeorgiou}, \citenamefont {Gupta}, \citenamefont
  {Senapati}, \citenamefont {Tiwari}, \citenamefont {Tai}, \citenamefont
  {Singh}, \citenamefont {Baldwin}, \citenamefont {Pfeiffer}, \citenamefont
  {Shayegan}, \citenamefont {Banerjee},\ and\ \citenamefont
  {Heiblum}}]{Alkalay26}%
  \BibitemOpen
  \bibfield  {author} {\bibinfo {author} {\bibfnamefont {T.}~\bibnamefont
  {Alkalay}}, \bibinfo {author} {\bibfnamefont {E.}~\bibnamefont
  {Hajigeorgiou}}, \bibinfo {author} {\bibfnamefont {A.}~\bibnamefont {Gupta}},
  \bibinfo {author} {\bibfnamefont {T.}~\bibnamefont {Senapati}}, \bibinfo
  {author} {\bibfnamefont {P.}~\bibnamefont {Tiwari}}, \bibinfo {author}
  {\bibfnamefont {C.-T.}\ \bibnamefont {Tai}}, \bibinfo {author} {\bibfnamefont
  {S.~K.}\ \bibnamefont {Singh}}, \bibinfo {author} {\bibfnamefont {K.~W.}\
  \bibnamefont {Baldwin}}, \bibinfo {author} {\bibfnamefont {L.~N.}\
  \bibnamefont {Pfeiffer}}, \bibinfo {author} {\bibfnamefont {M.}~\bibnamefont
  {Shayegan}}, \bibinfo {author} {\bibfnamefont {M.}~\bibnamefont {Banerjee}},\
  and\ \bibinfo {author} {\bibfnamefont {M.}~\bibnamefont {Heiblum}},\
  }\bibfield  {title} {\bibinfo {title} {Observation of $e/4$ charge at
  $\nu=1/2$ in a wide {GaAs} quantum well},\ }\href
  {https://doi.org/10.1103/c73x-q4z7} {\bibfield  {journal} {\bibinfo
  {journal} {Phys. Rev. Lett.}\ }\textbf {\bibinfo {volume} {137}},\ \bibinfo
  {pages} {136503} (\bibinfo {year} {2026})}\BibitemShut {NoStop}%
\bibitem [{\citenamefont {Feldman}\ and\ \citenamefont
  {Halperin}(2021)}]{Feldman21}%
  \BibitemOpen
  \bibfield  {author} {\bibinfo {author} {\bibfnamefont {D.~E.}\ \bibnamefont
  {Feldman}}\ and\ \bibinfo {author} {\bibfnamefont {B.~I.}\ \bibnamefont
  {Halperin}},\ }\bibfield  {title} {\bibinfo {title} {Fractional charge and
  fractional statistics in the quantum hall effects},\ }\href@noop {}
  {\bibfield  {journal} {\bibinfo  {journal} {Reports on Progress in Physics}\
  }\textbf {\bibinfo {volume} {84}},\ \bibinfo {pages} {076501} (\bibinfo
  {year} {2021})}\BibitemShut {NoStop}%
\bibitem [{\citenamefont {Faugno}\ \emph {et~al.}(2020)\citenamefont {Faugno},
  \citenamefont {Jain},\ and\ \citenamefont {Balram}}]{Faugno20A}%
  \BibitemOpen
  \bibfield  {author} {\bibinfo {author} {\bibfnamefont {W.~N.}\ \bibnamefont
  {Faugno}}, \bibinfo {author} {\bibfnamefont {J.~K.}\ \bibnamefont {Jain}},\
  and\ \bibinfo {author} {\bibfnamefont {A.~C.}\ \bibnamefont {Balram}},\
  }\bibfield  {title} {\bibinfo {title} {Non-abelian fractional quantum {Hall}
  state at $3/7$-filled {Landau} level},\ }\href
  {https://doi.org/10.1103/PhysRevResearch.2.033223} {\bibfield  {journal}
  {\bibinfo  {journal} {Phys. Rev. Research}\ }\textbf {\bibinfo {volume}
  {2}},\ \bibinfo {pages} {033223} (\bibinfo {year} {2020})}\BibitemShut
  {NoStop}%
\bibitem [{\citenamefont {Faugno}\ \emph {et~al.}(2021)\citenamefont {Faugno},
  \citenamefont {Zhao}, \citenamefont {Balram}, \citenamefont {Jolicoeur},\
  and\ \citenamefont {Jain}}]{Faugno21}%
  \BibitemOpen
  \bibfield  {author} {\bibinfo {author} {\bibfnamefont {W.~N.}\ \bibnamefont
  {Faugno}}, \bibinfo {author} {\bibfnamefont {T.}~\bibnamefont {Zhao}},
  \bibinfo {author} {\bibfnamefont {A.~C.}\ \bibnamefont {Balram}}, \bibinfo
  {author} {\bibfnamefont {T.}~\bibnamefont {Jolicoeur}},\ and\ \bibinfo
  {author} {\bibfnamefont {J.~K.}\ \bibnamefont {Jain}},\ }\bibfield  {title}
  {\bibinfo {title} {Unconventional ${\mathbb{z}}_{n}$ parton states at
  $\ensuremath{\nu}=7/3$: {Role} of finite width},\ }\href
  {https://doi.org/10.1103/PhysRevB.103.085303} {\bibfield  {journal} {\bibinfo
   {journal} {Phys. Rev. B}\ }\textbf {\bibinfo {volume} {103}},\ \bibinfo
  {pages} {085303} (\bibinfo {year} {2021})}\BibitemShut {NoStop}%
\bibitem [{\citenamefont {Gattu}\ and\ \citenamefont
  {Jain}(2025{\natexlab{a}})}]{Gattu25A}%
  \BibitemOpen
  \bibfield  {author} {\bibinfo {author} {\bibfnamefont {M.}~\bibnamefont
  {Gattu}}\ and\ \bibinfo {author} {\bibfnamefont {J.~K.}\ \bibnamefont
  {Jain}},\ }\bibfield  {title} {\bibinfo {title} {Molecular anyons in the
  fractional quantum {Hall} effect},\ }\href
  {https://doi.org/10.1103/scl5-8pv6} {\bibfield  {journal} {\bibinfo
  {journal} {Phys. Rev. Lett.}\ }\textbf {\bibinfo {volume} {135}},\ \bibinfo
  {pages} {236601} (\bibinfo {year} {2025}{\natexlab{a}})}\BibitemShut
  {NoStop}%
\bibitem [{\citenamefont {Xu}\ \emph {et~al.}(2025)\citenamefont {Xu},
  \citenamefont {Ji}, \citenamefont {Wang}, \citenamefont {Trung},\ and\
  \citenamefont {Yang}}]{Xu25}%
  \BibitemOpen
  \bibfield  {author} {\bibinfo {author} {\bibfnamefont {Q.}~\bibnamefont
  {Xu}}, \bibinfo {author} {\bibfnamefont {G.}~\bibnamefont {Ji}}, \bibinfo
  {author} {\bibfnamefont {Y.}~\bibnamefont {Wang}}, \bibinfo {author}
  {\bibfnamefont {H.~Q.}\ \bibnamefont {Trung}},\ and\ \bibinfo {author}
  {\bibfnamefont {B.}~\bibnamefont {Yang}},\ }\bibfield  {title} {\bibinfo
  {title} {Dynamics of anyon clusters in fractional quantum {Hall} fluids},\
  }\href {https://doi.org/10.1103/vgz6-z98r} {\bibfield  {journal} {\bibinfo
  {journal} {Phys. Rev. B}\ }\textbf {\bibinfo {volume} {112}},\ \bibinfo
  {pages} {235112} (\bibinfo {year} {2025})}\BibitemShut {NoStop}%
\bibitem [{\citenamefont {Wang}\ and\ \citenamefont {Zaletel}(2026)}]{Wang26A}%
  \BibitemOpen
  \bibfield  {author} {\bibinfo {author} {\bibfnamefont {T.}~\bibnamefont
  {Wang}}\ and\ \bibinfo {author} {\bibfnamefont {M.~P.}\ \bibnamefont
  {Zaletel}},\ }\bibfield  {title} {\bibinfo {title} {Anyon molecules in
  fractional quantum {Hall} states},\ }\href@noop {} {\bibfield  {journal}
  {\bibinfo  {journal} {arXiv:2604.09798}\ } (\bibinfo {year}
  {2026})}\BibitemShut {NoStop}%
\bibitem [{\citenamefont {Li}\ \emph {et~al.}(2026)\citenamefont {Li},
  \citenamefont {Nosov}, \citenamefont {Wang},\ and\ \citenamefont
  {Khalaf}}]{Li26}%
  \BibitemOpen
  \bibfield  {author} {\bibinfo {author} {\bibfnamefont {Q.}~\bibnamefont
  {Li}}, \bibinfo {author} {\bibfnamefont {P.~A.}\ \bibnamefont {Nosov}},
  \bibinfo {author} {\bibfnamefont {T.}~\bibnamefont {Wang}},\ and\ \bibinfo
  {author} {\bibfnamefont {E.}~\bibnamefont {Khalaf}},\ }\bibfield  {title}
  {\bibinfo {title} {Bound states of anyons: A geometric quantization
  approach},\ }\href@noop {} {\bibfield  {journal} {\bibinfo  {journal}
  {arXiv:2603.24701}\ } (\bibinfo {year} {2026})}\BibitemShut {NoStop}%
\bibitem [{\citenamefont {Jain}(1989)}]{Jain89B}%
  \BibitemOpen
  \bibfield  {author} {\bibinfo {author} {\bibfnamefont {J.~K.}\ \bibnamefont
  {Jain}},\ }\bibfield  {title} {\bibinfo {title} {Incompressible quantum
  {Hall} states},\ }\href {https://doi.org/10.1103/PhysRevB.40.8079} {\bibfield
   {journal} {\bibinfo  {journal} {Phys. Rev. B}\ }\textbf {\bibinfo {volume}
  {40}},\ \bibinfo {pages} {8079} (\bibinfo {year} {1989})}\BibitemShut
  {NoStop}%
\bibitem [{\citenamefont {Wen}(1991)}]{Wen91}%
  \BibitemOpen
  \bibfield  {author} {\bibinfo {author} {\bibfnamefont {X.~G.}\ \bibnamefont
  {Wen}},\ }\bibfield  {title} {\bibinfo {title} {Non-abelian statistics in the
  fractional quantum {Hall} states},\ }\href
  {https://doi.org/10.1103/PhysRevLett.66.802} {\bibfield  {journal} {\bibinfo
  {journal} {Phys. Rev. Lett.}\ }\textbf {\bibinfo {volume} {66}},\ \bibinfo
  {pages} {802} (\bibinfo {year} {1991})}\BibitemShut {NoStop}%
\bibitem [{\citenamefont {Haldane}(1983)}]{Haldane83}%
  \BibitemOpen
  \bibfield  {author} {\bibinfo {author} {\bibfnamefont {F.~D.~M.}\
  \bibnamefont {Haldane}},\ }\bibfield  {title} {\bibinfo {title} {Fractional
  quantization of the {Hall} effect: A hierarchy of incompressible quantum
  fluid states},\ }\href {https://doi.org/10.1103/PhysRevLett.51.605}
  {\bibfield  {journal} {\bibinfo  {journal} {Phys. Rev. Lett.}\ }\textbf
  {\bibinfo {volume} {51}},\ \bibinfo {pages} {605} (\bibinfo {year}
  {1983})}\BibitemShut {NoStop}%
\bibitem [{\citenamefont {Wu}\ and\ \citenamefont {Yang}(1976)}]{Wu76}%
  \BibitemOpen
  \bibfield  {author} {\bibinfo {author} {\bibfnamefont {T.~T.}\ \bibnamefont
  {Wu}}\ and\ \bibinfo {author} {\bibfnamefont {C.~N.}\ \bibnamefont {Yang}},\
  }\bibfield  {title} {\bibinfo {title} {Dirac monopole without strings:
  Monopole harmonics},\ }\href {https://doi.org/10.1016/0550-3213(76)90143-7}
  {\bibfield  {journal} {\bibinfo  {journal} {Nucl. Phys. B}\ }\textbf
  {\bibinfo {volume} {107}},\ \bibinfo {pages} {365} (\bibinfo {year}
  {1976})}\BibitemShut {NoStop}%
\bibitem [{\citenamefont {Wu}\ and\ \citenamefont {Yang}(1977)}]{Wu77}%
  \BibitemOpen
  \bibfield  {author} {\bibinfo {author} {\bibfnamefont {T.~T.}\ \bibnamefont
  {Wu}}\ and\ \bibinfo {author} {\bibfnamefont {C.~N.}\ \bibnamefont {Yang}},\
  }\bibfield  {title} {\bibinfo {title} {Some properties of monopole
  harmonics},\ }\href {https://doi.org/10.1103/physrevd.16.1018} {\bibfield
  {journal} {\bibinfo  {journal} {Phys. Rev. D}\ }\textbf {\bibinfo {volume}
  {16}},\ \bibinfo {pages} {1018} (\bibinfo {year} {1977})}\BibitemShut
  {NoStop}%
\bibitem [{SM()}]{SM}%
  \BibitemOpen
  \href@noop {} {}\bibinfo {howpublished} {See Supplemental Material appended below for (i) a review of the spherical geometry,
  (ii) the parton wave functions in this geometry, (iii) the construction of
  the two-quasihole and two-quasiparticle states, (iv) the effective
  interactions used for wide quantum wells and for higher Landau levels of GaAs
  and monolayer graphene, and (v) additional results. It also includes
  additional
  references~\cite{Jain07,Varshalovich88,Yutushui25,CastroNeto09,Goerbig11,Balram15D}}\BibitemShut
  {NoStop}%
\bibitem [{\citenamefont {Jain}\ and\ \citenamefont
  {Kamilla}(1997{\natexlab{a}})}]{Jain97A}%
  \BibitemOpen
  \bibfield  {author} {\bibinfo {author} {\bibfnamefont {J.~K.}\ \bibnamefont
  {Jain}}\ and\ \bibinfo {author} {\bibfnamefont {R.~K.}\ \bibnamefont
  {Kamilla}},\ }\bibfield  {title} {\bibinfo {title} {Composite fermions in the
  {Hilbert} space of the lowest electronic {Landau} level},\ }\href
  {https://doi.org/10.1142/S0217979297001301} {\bibfield  {journal} {\bibinfo
  {journal} {Int. J. Mod. Phys. B}\ }\textbf {\bibinfo {volume} {11}},\
  \bibinfo {pages} {2621} (\bibinfo {year} {1997}{\natexlab{a}})}\BibitemShut
  {NoStop}%
\bibitem [{\citenamefont {Jain}\ and\ \citenamefont
  {Kamilla}(1997{\natexlab{b}})}]{Jain97}%
  \BibitemOpen
  \bibfield  {author} {\bibinfo {author} {\bibfnamefont {J.~K.}\ \bibnamefont
  {Jain}}\ and\ \bibinfo {author} {\bibfnamefont {R.~K.}\ \bibnamefont
  {Kamilla}},\ }\bibfield  {title} {\bibinfo {title} {Quantitative study of
  large composite-fermion systems},\ }\href
  {https://doi.org/10.1103/PhysRevB.55.R4895} {\bibfield  {journal} {\bibinfo
  {journal} {Phys. Rev. B}\ }\textbf {\bibinfo {volume} {55}},\ \bibinfo
  {pages} {R4895} (\bibinfo {year} {1997}{\natexlab{b}})}\BibitemShut {NoStop}%
\bibitem [{\citenamefont {Gattu}\ and\ \citenamefont
  {Jain}(2025{\natexlab{b}})}]{Gattu25}%
  \BibitemOpen
  \bibfield  {author} {\bibinfo {author} {\bibfnamefont {M.}~\bibnamefont
  {Gattu}}\ and\ \bibinfo {author} {\bibfnamefont {J.~K.}\ \bibnamefont
  {Jain}},\ }\bibfield  {title} {\bibinfo {title} {Unlocking new regimes in
  fractional quantum hall effect with quaternions},\ }\href
  {https://doi.org/10.1103/physrevlett.134.156501} {\bibfield  {journal}
  {\bibinfo  {journal} {Phys. Rev. Lett.}\ }\textbf {\bibinfo {volume} {134}},\
  \bibinfo {pages} {156501} (\bibinfo {year} {2025}{\natexlab{b}})}\BibitemShut
  {NoStop}%
\bibitem [{\citenamefont {Wu}\ \emph {et~al.}(2017)\citenamefont {Wu},
  \citenamefont {Shi},\ and\ \citenamefont {Jain}}]{Wu17}%
  \BibitemOpen
  \bibfield  {author} {\bibinfo {author} {\bibfnamefont {Y.-H.}\ \bibnamefont
  {Wu}}, \bibinfo {author} {\bibfnamefont {T.}~\bibnamefont {Shi}},\ and\
  \bibinfo {author} {\bibfnamefont {J.~K.}\ \bibnamefont {Jain}},\ }\bibfield
  {title} {\bibinfo {title} {Non-abelian parton fractional quantum hall effect
  in multilayer graphene},\ }\href@noop {} {\bibfield  {journal} {\bibinfo
  {journal} {Nano Letters}\ }\textbf {\bibinfo {volume} {17}},\ \bibinfo
  {pages} {4643} (\bibinfo {year} {2017})}\BibitemShut {NoStop}%
\bibitem [{\citenamefont {Efron}(1982)}]{Efron82}%
  \BibitemOpen
  \bibfield  {author} {\bibinfo {author} {\bibfnamefont {B.}~\bibnamefont
  {Efron}},\ }\href {https://doi.org/10.1137/1.9781611970319} {\emph {\bibinfo
  {title} {The Jackknife, the Bootstrap and Other Resampling Plans}}}\
  (\bibinfo  {publisher} {Society for Industrial and Applied Mathematics},\
  \bibinfo {address} {Philadelphia},\ \bibinfo {year} {1982})\BibitemShut
  {NoStop}%
\bibitem [{\citenamefont {M\"oller}\ \emph {et~al.}(2011)\citenamefont
  {M\"oller}, \citenamefont {W\'ojs},\ and\ \citenamefont {Cooper}}]{Moller11}%
  \BibitemOpen
  \bibfield  {author} {\bibinfo {author} {\bibfnamefont {G.}~\bibnamefont
  {M\"oller}}, \bibinfo {author} {\bibfnamefont {A.}~\bibnamefont {W\'ojs}},\
  and\ \bibinfo {author} {\bibfnamefont {N.~R.}\ \bibnamefont {Cooper}},\
  }\bibfield  {title} {\bibinfo {title} {Neutral fermion excitations in the
  {Moore}-{Read} state at filling factor $\ensuremath{\nu}=5/2$},\ }\href
  {https://doi.org/10.1103/PhysRevLett.107.036803} {\bibfield  {journal}
  {\bibinfo  {journal} {Phys. Rev. Lett.}\ }\textbf {\bibinfo {volume} {107}},\
  \bibinfo {pages} {036803} (\bibinfo {year} {2011})}\BibitemShut {NoStop}%
\bibitem [{\citenamefont {Sreejith}\ \emph {et~al.}(2011)\citenamefont
  {Sreejith}, \citenamefont {T\H{o}ke}, \citenamefont {W\'ojs},\ and\
  \citenamefont {Jain}}]{Sreejith11}%
  \BibitemOpen
  \bibfield  {author} {\bibinfo {author} {\bibfnamefont {G.~J.}\ \bibnamefont
  {Sreejith}}, \bibinfo {author} {\bibfnamefont {C.}~\bibnamefont {T\H{o}ke}},
  \bibinfo {author} {\bibfnamefont {A.}~\bibnamefont {W\'ojs}},\ and\ \bibinfo
  {author} {\bibfnamefont {J.~K.}\ \bibnamefont {Jain}},\ }\bibfield  {title}
  {\bibinfo {title} {Bipartite composite fermion states},\ }\href
  {https://doi.org/10.1103/PhysRevLett.107.086806} {\bibfield  {journal}
  {\bibinfo  {journal} {Phys. Rev. Lett.}\ }\textbf {\bibinfo {volume} {107}},\
  \bibinfo {pages} {086806} (\bibinfo {year} {2011})}\BibitemShut {NoStop}%
\bibitem [{\citenamefont {Pu}\ \emph {et~al.}(2023)\citenamefont {Pu},
  \citenamefont {Balram}, \citenamefont {Fremling}, \citenamefont {Gromov},\
  and\ \citenamefont {Papi\ifmmode~\acute{c}\else \'{c}\fi{}}}]{Pu23}%
  \BibitemOpen
  \bibfield  {author} {\bibinfo {author} {\bibfnamefont {S.}~\bibnamefont
  {Pu}}, \bibinfo {author} {\bibfnamefont {A.~C.}\ \bibnamefont {Balram}},
  \bibinfo {author} {\bibfnamefont {M.}~\bibnamefont {Fremling}}, \bibinfo
  {author} {\bibfnamefont {A.}~\bibnamefont {Gromov}},\ and\ \bibinfo {author}
  {\bibfnamefont {Z.}~\bibnamefont {Papi\ifmmode~\acute{c}\else \'{c}\fi{}}},\
  }\bibfield  {title} {\bibinfo {title} {Signatures of supersymmetry in the
  $\ensuremath{\nu}=5/2$ fractional quantum {Hall} effect},\ }\href
  {https://doi.org/10.1103/PhysRevLett.130.176501} {\bibfield  {journal}
  {\bibinfo  {journal} {Phys. Rev. Lett.}\ }\textbf {\bibinfo {volume} {130}},\
  \bibinfo {pages} {176501} (\bibinfo {year} {2023})}\BibitemShut {NoStop}%
\bibitem [{\citenamefont {Suen}\ \emph {et~al.}(1994)\citenamefont {Suen},
  \citenamefont {Manoharan}, \citenamefont {Ying}, \citenamefont {Santos},\
  and\ \citenamefont {Shayegan}}]{Suen94A}%
  \BibitemOpen
  \bibfield  {author} {\bibinfo {author} {\bibfnamefont {Y.}~\bibnamefont
  {Suen}}, \bibinfo {author} {\bibfnamefont {H.}~\bibnamefont {Manoharan}},
  \bibinfo {author} {\bibfnamefont {X.}~\bibnamefont {Ying}}, \bibinfo {author}
  {\bibfnamefont {M.}~\bibnamefont {Santos}},\ and\ \bibinfo {author}
  {\bibfnamefont {M.}~\bibnamefont {Shayegan}},\ }\bibfield  {title} {\bibinfo
  {title} {One-component to two-component transitions of fractional quantum
  {Hall} states in a wide quantum well},\ }\href
  {https://doi.org/https://doi.org/10.1016/0039-6028(94)90852-4} {\bibfield
  {journal} {\bibinfo  {journal} {Surface Science}\ }\textbf {\bibinfo {volume}
  {305}},\ \bibinfo {pages} {13 } (\bibinfo {year} {1994})}\BibitemShut
  {NoStop}%
\bibitem [{\citenamefont {Tsui}\ \emph {et~al.}(2024)\citenamefont {Tsui},
  \citenamefont {He}, \citenamefont {Hu}, \citenamefont {Lake}, \citenamefont
  {Wang}, \citenamefont {Watanabe}, \citenamefont {Taniguchi}, \citenamefont
  {Zaletel},\ and\ \citenamefont {Yazdani}}]{Tsui24}%
  \BibitemOpen
  \bibfield  {author} {\bibinfo {author} {\bibfnamefont {Y.-C.}\ \bibnamefont
  {Tsui}}, \bibinfo {author} {\bibfnamefont {M.}~\bibnamefont {He}}, \bibinfo
  {author} {\bibfnamefont {Y.}~\bibnamefont {Hu}}, \bibinfo {author}
  {\bibfnamefont {E.}~\bibnamefont {Lake}}, \bibinfo {author} {\bibfnamefont
  {T.}~\bibnamefont {Wang}}, \bibinfo {author} {\bibfnamefont {K.}~\bibnamefont
  {Watanabe}}, \bibinfo {author} {\bibfnamefont {T.}~\bibnamefont {Taniguchi}},
  \bibinfo {author} {\bibfnamefont {M.~P.}\ \bibnamefont {Zaletel}},\ and\
  \bibinfo {author} {\bibfnamefont {A.}~\bibnamefont {Yazdani}},\ }\bibfield
  {title} {\bibinfo {title} {Direct observation of a magnetic-field-induced
  wigner crystal},\ }\href {https://doi.org/10.1038/s41586-024-07212-7}
  {\bibfield  {journal} {\bibinfo  {journal} {Nature}\ }\textbf {\bibinfo
  {volume} {628}},\ \bibinfo {pages} {287–292} (\bibinfo {year}
  {2024})}\BibitemShut {NoStop}%
\bibitem [{\citenamefont {Chiu}\ \emph {et~al.}(2025)\citenamefont {Chiu},
  \citenamefont {Wang}, \citenamefont {Fan}, \citenamefont {Watanabe},
  \citenamefont {Taniguchi}, \citenamefont {Liu}, \citenamefont {Zaletel},\
  and\ \citenamefont {Yazdani}}]{Chiu25}%
  \BibitemOpen
  \bibfield  {author} {\bibinfo {author} {\bibfnamefont {C.-L.}\ \bibnamefont
  {Chiu}}, \bibinfo {author} {\bibfnamefont {T.}~\bibnamefont {Wang}}, \bibinfo
  {author} {\bibfnamefont {R.}~\bibnamefont {Fan}}, \bibinfo {author}
  {\bibfnamefont {K.}~\bibnamefont {Watanabe}}, \bibinfo {author}
  {\bibfnamefont {T.}~\bibnamefont {Taniguchi}}, \bibinfo {author}
  {\bibfnamefont {X.}~\bibnamefont {Liu}}, \bibinfo {author} {\bibfnamefont
  {M.~P.}\ \bibnamefont {Zaletel}},\ and\ \bibinfo {author} {\bibfnamefont
  {A.}~\bibnamefont {Yazdani}},\ }\bibfield  {title} {\bibinfo {title} {High
  spatial resolution charge sensing of quantum hall states},\ }\href
  {https://doi.org/10.1073/pnas.2424781122} {\bibfield  {journal} {\bibinfo
  {journal} {Proceedings of the National Academy of Sciences}\ }\textbf
  {\bibinfo {volume} {122}},\ \bibinfo {pages} {e2424781122} (\bibinfo {year}
  {2025})}\BibitemShut {NoStop}%
\bibitem [{\citenamefont {Yang}\ and\ \citenamefont {Halperin}(2009)}]{Yang09}%
  \BibitemOpen
  \bibfield  {author} {\bibinfo {author} {\bibfnamefont {K.}~\bibnamefont
  {Yang}}\ and\ \bibinfo {author} {\bibfnamefont {B.~I.}\ \bibnamefont
  {Halperin}},\ }\bibfield  {title} {\bibinfo {title} {Thermopower as a
  possible probe of non-abelian quasiparticle statistics in fractional quantum
  hall liquids},\ }\href {https://doi.org/10.1103/PhysRevB.79.115317}
  {\bibfield  {journal} {\bibinfo  {journal} {Phys. Rev. B}\ }\textbf {\bibinfo
  {volume} {79}},\ \bibinfo {pages} {115317} (\bibinfo {year}
  {2009})}\BibitemShut {NoStop}%
\bibitem [{\citenamefont {Melcer}\ \emph {et~al.}(2024)\citenamefont {Melcer},
  \citenamefont {Gil}, \citenamefont {Paul}, \citenamefont {Tiwari},
  \citenamefont {Umansky}, \citenamefont {Heiblum}, \citenamefont {Oreg},
  \citenamefont {Stern},\ and\ \citenamefont {Berg}}]{Melcer2024heat}%
  \BibitemOpen
  \bibfield  {author} {\bibinfo {author} {\bibfnamefont {R.~A.}\ \bibnamefont
  {Melcer}}, \bibinfo {author} {\bibfnamefont {A.}~\bibnamefont {Gil}},
  \bibinfo {author} {\bibfnamefont {A.~K.}\ \bibnamefont {Paul}}, \bibinfo
  {author} {\bibfnamefont {P.}~\bibnamefont {Tiwari}}, \bibinfo {author}
  {\bibfnamefont {V.}~\bibnamefont {Umansky}}, \bibinfo {author} {\bibfnamefont
  {M.}~\bibnamefont {Heiblum}}, \bibinfo {author} {\bibfnamefont
  {Y.}~\bibnamefont {Oreg}}, \bibinfo {author} {\bibfnamefont {A.}~\bibnamefont
  {Stern}},\ and\ \bibinfo {author} {\bibfnamefont {E.}~\bibnamefont {Berg}},\
  }\bibfield  {title} {\bibinfo {title} {Heat conductance of the quantum hall
  bulk},\ }\href@noop {} {\bibfield  {journal} {\bibinfo  {journal} {Nature}\
  }\textbf {\bibinfo {volume} {625}},\ \bibinfo {pages} {489} (\bibinfo {year}
  {2024})}\BibitemShut {NoStop}%
\bibitem [{\citenamefont {Boivin}\ \emph {et~al.}(2026)\citenamefont {Boivin},
  \citenamefont {Petrescu}, \citenamefont {Berkson-Korenberg}, \citenamefont
  {West}, \citenamefont {Pfeiffer},\ and\ \citenamefont {Gervais}}]{Boivin26}%
  \BibitemOpen
  \bibfield  {author} {\bibinfo {author} {\bibfnamefont {F.}~\bibnamefont
  {Boivin}}, \bibinfo {author} {\bibfnamefont {M.}~\bibnamefont {Petrescu}},
  \bibinfo {author} {\bibfnamefont {Z.}~\bibnamefont {Berkson-Korenberg}},
  \bibinfo {author} {\bibfnamefont {K.~W.}\ \bibnamefont {West}}, \bibinfo
  {author} {\bibfnamefont {L.~N.}\ \bibnamefont {Pfeiffer}},\ and\ \bibinfo
  {author} {\bibfnamefont {G.}~\bibnamefont {Gervais}},\ }\bibfield  {title}
  {\bibinfo {title} {Bulk thermal conductance of the 5/2 and 7/3 fractional
  quantum hall states in the corbino geometry},\ }\href
  {https://doi.org/10.1103/3mm9-47h9} {\bibfield  {journal} {\bibinfo
  {journal} {Phys. Rev. B}\ }\textbf {\bibinfo {volume} {113}},\ \bibinfo
  {pages} {L201301} (\bibinfo {year} {2026})}\BibitemShut {NoStop}%
\bibitem [{\citenamefont {Bonderson}\ \emph
  {et~al.}(2006{\natexlab{a}})\citenamefont {Bonderson}, \citenamefont
  {Kitaev},\ and\ \citenamefont {Shtengel}}]{Bonderson06}%
  \BibitemOpen
  \bibfield  {author} {\bibinfo {author} {\bibfnamefont {P.}~\bibnamefont
  {Bonderson}}, \bibinfo {author} {\bibfnamefont {A.}~\bibnamefont {Kitaev}},\
  and\ \bibinfo {author} {\bibfnamefont {K.}~\bibnamefont {Shtengel}},\
  }\bibfield  {title} {\bibinfo {title} {Detecting non-{Abelian} statistics in
  the $\nu=5/2$ fractional quantum {Hall} state},\ }\href
  {https://doi.org/10.1103/PhysRevLett.96.016803} {\bibfield  {journal}
  {\bibinfo  {journal} {Phys. Rev. Lett.}\ }\textbf {\bibinfo {volume} {96}},\
  \bibinfo {pages} {016803} (\bibinfo {year} {2006}{\natexlab{a}})}\BibitemShut
  {NoStop}%
\bibitem [{\citenamefont {Bonderson}\ \emph
  {et~al.}(2006{\natexlab{b}})\citenamefont {Bonderson}, \citenamefont
  {Shtengel},\ and\ \citenamefont {Slingerland}}]{Bonderson06A}%
  \BibitemOpen
  \bibfield  {author} {\bibinfo {author} {\bibfnamefont {P.}~\bibnamefont
  {Bonderson}}, \bibinfo {author} {\bibfnamefont {K.}~\bibnamefont
  {Shtengel}},\ and\ \bibinfo {author} {\bibfnamefont {J.~K.}\ \bibnamefont
  {Slingerland}},\ }\bibfield  {title} {\bibinfo {title} {Probing non-{Abelian}
  statistics with quasiparticle interferometry},\ }\href
  {https://doi.org/10.1103/PhysRevLett.97.016401} {\bibfield  {journal}
  {\bibinfo  {journal} {Phys. Rev. Lett.}\ }\textbf {\bibinfo {volume} {97}},\
  \bibinfo {pages} {016401} (\bibinfo {year} {2006}{\natexlab{b}})}\BibitemShut
  {NoStop}%
\bibitem [{\citenamefont {Willett}\ \emph {et~al.}(2009)\citenamefont
  {Willett}, \citenamefont {Pfeiffer},\ and\ \citenamefont {West}}]{Willett09}%
  \BibitemOpen
  \bibfield  {author} {\bibinfo {author} {\bibfnamefont {R.~L.}\ \bibnamefont
  {Willett}}, \bibinfo {author} {\bibfnamefont {L.~N.}\ \bibnamefont
  {Pfeiffer}},\ and\ \bibinfo {author} {\bibfnamefont {K.~W.}\ \bibnamefont
  {West}},\ }\bibfield  {title} {\bibinfo {title} {Measurement of filling
  factor 5/2 quasiparticle interference with observation of charge e/4 and e/2
  period oscillations},\ }\href {https://doi.org/10.1073/pnas.0812599106}
  {\bibfield  {journal} {\bibinfo  {journal} {Proc. Natl. Acad. Sci. USA}\
  }\textbf {\bibinfo {volume} {106}},\ \bibinfo {pages} {8853} (\bibinfo {year}
  {2009})}\BibitemShut {NoStop}%
\bibitem [{\citenamefont {Bishara}\ and\ \citenamefont
  {Nayak}(2009)}]{Bishara09}%
  \BibitemOpen
  \bibfield  {author} {\bibinfo {author} {\bibfnamefont {W.}~\bibnamefont
  {Bishara}}\ and\ \bibinfo {author} {\bibfnamefont {C.}~\bibnamefont
  {Nayak}},\ }\bibfield  {title} {\bibinfo {title} {Effect of {Landau} level
  mixing on the effective interaction between electrons in the fractional
  quantum {Hall} regime},\ }\href {https://doi.org/10.1103/PhysRevB.80.121302}
  {\bibfield  {journal} {\bibinfo  {journal} {Phys. Rev. B}\ }\textbf {\bibinfo
  {volume} {80}},\ \bibinfo {pages} {121302} (\bibinfo {year}
  {2009})}\BibitemShut {NoStop}%
\bibitem [{\citenamefont {Rezayi}\ and\ \citenamefont
  {Simon}(2011)}]{Rezayi11}%
  \BibitemOpen
  \bibfield  {author} {\bibinfo {author} {\bibfnamefont {E.~H.}\ \bibnamefont
  {Rezayi}}\ and\ \bibinfo {author} {\bibfnamefont {S.~H.}\ \bibnamefont
  {Simon}},\ }\bibfield  {title} {\bibinfo {title} {Breaking of particle-hole
  symmetry by {Landau} level mixing in the $\nu=5/2$ quantized {Hall} state},\
  }\href {https://doi.org/10.1103/PhysRevLett.106.116801} {\bibfield  {journal}
  {\bibinfo  {journal} {Phys. Rev. Lett.}\ }\textbf {\bibinfo {volume} {106}},\
  \bibinfo {pages} {116801} (\bibinfo {year} {2011})}\BibitemShut {NoStop}%
\bibitem [{\citenamefont {Peterson}\ and\ \citenamefont
  {Nayak}(2013)}]{Peterson13}%
  \BibitemOpen
  \bibfield  {author} {\bibinfo {author} {\bibfnamefont {M.~R.}\ \bibnamefont
  {Peterson}}\ and\ \bibinfo {author} {\bibfnamefont {C.}~\bibnamefont
  {Nayak}},\ }\bibfield  {title} {\bibinfo {title} {More realistic
  {Hamiltonians} for the fractional quantum {Hall} regime in {Ga}{As} and
  graphene},\ }\href {https://doi.org/10.1103/PhysRevB.87.245129} {\bibfield
  {journal} {\bibinfo  {journal} {Phys. Rev. B}\ }\textbf {\bibinfo {volume}
  {87}},\ \bibinfo {pages} {245129} (\bibinfo {year} {2013})}\BibitemShut
  {NoStop}%
\bibitem [{\citenamefont {Sodemann}\ and\ \citenamefont
  {MacDonald}(2013)}]{Sodemann13}%
  \BibitemOpen
  \bibfield  {author} {\bibinfo {author} {\bibfnamefont {I.}~\bibnamefont
  {Sodemann}}\ and\ \bibinfo {author} {\bibfnamefont {A.~H.}\ \bibnamefont
  {MacDonald}},\ }\bibfield  {title} {\bibinfo {title} {{Landau} level mixing
  and the fractional quantum {Hall} effect},\ }\href
  {https://doi.org/10.1103/PhysRevB.87.245425} {\bibfield  {journal} {\bibinfo
  {journal} {Phys. Rev. B}\ }\textbf {\bibinfo {volume} {87}},\ \bibinfo
  {pages} {245425} (\bibinfo {year} {2013})}\BibitemShut {NoStop}%
\bibitem [{\citenamefont {Pakrouski}\ \emph {et~al.}(2015)\citenamefont
  {Pakrouski}, \citenamefont {Peterson}, \citenamefont {Jolicoeur},
  \citenamefont {Scarola}, \citenamefont {Nayak},\ and\ \citenamefont
  {Troyer}}]{Pakrouski15}%
  \BibitemOpen
  \bibfield  {author} {\bibinfo {author} {\bibfnamefont {K.}~\bibnamefont
  {Pakrouski}}, \bibinfo {author} {\bibfnamefont {M.~R.}\ \bibnamefont
  {Peterson}}, \bibinfo {author} {\bibfnamefont {T.}~\bibnamefont {Jolicoeur}},
  \bibinfo {author} {\bibfnamefont {V.~W.}\ \bibnamefont {Scarola}}, \bibinfo
  {author} {\bibfnamefont {C.}~\bibnamefont {Nayak}},\ and\ \bibinfo {author}
  {\bibfnamefont {M.}~\bibnamefont {Troyer}},\ }\bibfield  {title} {\bibinfo
  {title} {Phase diagram of the $\nu=5/2$ fractional quantum {Hall} effect:
  Effects of {Landau}-level mixing and nonzero width},\ }\href
  {https://doi.org/10.1103/PhysRevX.5.021004} {\bibfield  {journal} {\bibinfo
  {journal} {Phys. Rev. X}\ }\textbf {\bibinfo {volume} {5}},\ \bibinfo {pages}
  {021004} (\bibinfo {year} {2015})}\BibitemShut {NoStop}%
\bibitem [{\citenamefont {Rezayi}(2017)}]{Rezayi17}%
  \BibitemOpen
  \bibfield  {author} {\bibinfo {author} {\bibfnamefont {E.~H.}\ \bibnamefont
  {Rezayi}},\ }\bibfield  {title} {\bibinfo {title} {Landau level mixing and
  the ground state of the $\ensuremath{\nu}=5/2$ quantum {Hall} effect},\
  }\href {https://doi.org/10.1103/PhysRevLett.119.026801} {\bibfield  {journal}
  {\bibinfo  {journal} {Phys. Rev. Lett.}\ }\textbf {\bibinfo {volume} {119}},\
  \bibinfo {pages} {026801} (\bibinfo {year} {2017})}\BibitemShut {NoStop}%
\bibitem [{\citenamefont {Jain}(2007)}]{Jain07}%
  \BibitemOpen
  \bibfield  {author} {\bibinfo {author} {\bibfnamefont {J.~K.}\ \bibnamefont
  {Jain}},\ }\href@noop {} {\emph {\bibinfo {title} {Composite Fermions}}}\
  (\bibinfo  {publisher} {Cambridge University Press, New York, US},\ \bibinfo
  {year} {2007})\BibitemShut {NoStop}%
\bibitem [{\citenamefont {Varshalovich}\ \emph {et~al.}(1988)\citenamefont
  {Varshalovich}, \citenamefont {Moskalev},\ and\ \citenamefont
  {Khersonskii}}]{Varshalovich88}%
  \BibitemOpen
  \bibfield  {author} {\bibinfo {author} {\bibfnamefont {D.~A.}\ \bibnamefont
  {Varshalovich}}, \bibinfo {author} {\bibfnamefont {A.~N.}\ \bibnamefont
  {Moskalev}},\ and\ \bibinfo {author} {\bibfnamefont {V.~K.}\ \bibnamefont
  {Khersonskii}},\ }\href {https://doi.org/10.1142/0270} {\emph {\bibinfo
  {title} {Quantum Theory of Angular Momentum}}}\ (\bibinfo  {publisher} {WORLD
  SCIENTIFIC},\ \bibinfo {year} {1988})\BibitemShut {NoStop}%
\bibitem [{\citenamefont {Yutushui}\ and\ \citenamefont
  {Mross}(2025)}]{Yutushui25}%
  \BibitemOpen
  \bibfield  {author} {\bibinfo {author} {\bibfnamefont {M.}~\bibnamefont
  {Yutushui}}\ and\ \bibinfo {author} {\bibfnamefont {D.~F.}\ \bibnamefont
  {Mross}},\ }\bibfield  {title} {\bibinfo {title} {Phase diagram of
  compressible and paired states in the quarter-filled landau level},\ }\href
  {https://doi.org/10.1103/PhysRevB.111.035106} {\bibfield  {journal} {\bibinfo
   {journal} {Phys. Rev. B}\ }\textbf {\bibinfo {volume} {111}},\ \bibinfo
  {pages} {035106} (\bibinfo {year} {2025})}\BibitemShut {NoStop}%
\bibitem [{\citenamefont {Castro~Neto}\ \emph {et~al.}(2009)\citenamefont
  {Castro~Neto}, \citenamefont {Guinea}, \citenamefont {Peres}, \citenamefont
  {Novoselov},\ and\ \citenamefont {Geim}}]{CastroNeto09}%
  \BibitemOpen
  \bibfield  {author} {\bibinfo {author} {\bibfnamefont {A.~H.}\ \bibnamefont
  {Castro~Neto}}, \bibinfo {author} {\bibfnamefont {F.}~\bibnamefont {Guinea}},
  \bibinfo {author} {\bibfnamefont {N.~M.~R.}\ \bibnamefont {Peres}}, \bibinfo
  {author} {\bibfnamefont {K.~S.}\ \bibnamefont {Novoselov}},\ and\ \bibinfo
  {author} {\bibfnamefont {A.~K.}\ \bibnamefont {Geim}},\ }\bibfield  {title}
  {\bibinfo {title} {The electronic properties of graphene},\ }\href
  {https://doi.org/10.1103/RevModPhys.81.109} {\bibfield  {journal} {\bibinfo
  {journal} {Rev. Mod. Phys.}\ }\textbf {\bibinfo {volume} {81}},\ \bibinfo
  {pages} {109} (\bibinfo {year} {2009})}\BibitemShut {NoStop}%
\bibitem [{\citenamefont {Goerbig}(2011)}]{Goerbig11}%
  \BibitemOpen
  \bibfield  {author} {\bibinfo {author} {\bibfnamefont {M.~O.}\ \bibnamefont
  {Goerbig}},\ }\bibfield  {title} {\bibinfo {title} {Electronic properties of
  graphene in a strong magnetic field},\ }\href
  {https://doi.org/10.1103/RevModPhys.83.1193} {\bibfield  {journal} {\bibinfo
  {journal} {Rev. Mod. Phys.}\ }\textbf {\bibinfo {volume} {83}},\ \bibinfo
  {pages} {1193} (\bibinfo {year} {2011})}\BibitemShut {NoStop}%
\bibitem [{\citenamefont {Balram}\ \emph {et~al.}(2015)\citenamefont {Balram},
  \citenamefont {T\ifmmode~\mbox{\H{o}}\else \H{o}\fi{}ke}, \citenamefont
  {W\'ojs},\ and\ \citenamefont {Jain}}]{Balram15D}%
  \BibitemOpen
  \bibfield  {author} {\bibinfo {author} {\bibfnamefont {A.~C.}\ \bibnamefont
  {Balram}}, \bibinfo {author} {\bibfnamefont {C.}~\bibnamefont
  {T\ifmmode~\mbox{\H{o}}\else \H{o}\fi{}ke}}, \bibinfo {author} {\bibfnamefont
  {A.}~\bibnamefont {W\'ojs}},\ and\ \bibinfo {author} {\bibfnamefont {J.~K.}\
  \bibnamefont {Jain}},\ }\bibfield  {title} {\bibinfo {title} {Spontaneous
  polarization of composite fermions in the $n=1$ {Landau} level of graphene},\
  }\href {https://doi.org/10.1103/PhysRevB.92.205120} {\bibfield  {journal}
  {\bibinfo  {journal} {Phys. Rev. B}\ }\textbf {\bibinfo {volume} {92}},\
  \bibinfo {pages} {205120} (\bibinfo {year} {2015})}\BibitemShut {NoStop}%
\bibitem [{\citenamefont {Ortalano}\ \emph {et~al.}(1997)\citenamefont
  {Ortalano}, \citenamefont {He},\ and\ \citenamefont
  {Das~Sarma}}]{Ortalano97}%
  \BibitemOpen
  \bibfield  {author} {\bibinfo {author} {\bibfnamefont {M.~W.}\ \bibnamefont
  {Ortalano}}, \bibinfo {author} {\bibfnamefont {S.}~\bibnamefont {He}},\ and\
  \bibinfo {author} {\bibfnamefont {S.}~\bibnamefont {Das~Sarma}},\ }\bibfield
  {title} {\bibinfo {title} {Realistic calculations of correlated
  incompressible electronic states in {GaAs-Al$_{x}$Ga$_{1-x}$As}
  heterostructures and quantum wells},\ }\href
  {https://doi.org/10.1103/PhysRevB.55.7702} {\bibfield  {journal} {\bibinfo
  {journal} {Phys. Rev. B}\ }\textbf {\bibinfo {volume} {55}},\ \bibinfo
  {pages} {7702} (\bibinfo {year} {1997})}\BibitemShut {NoStop}%
\bibitem [{\citenamefont {Rother}(2020)}]{Rother09}%
  \BibitemOpen
  \bibfield  {author} {\bibinfo {author} {\bibfnamefont {M.}~\bibnamefont
  {Rother}},\ }\href@noop {} {\bibinfo {title} {2{D} {Schroedinger} {Poisson}
  solver {AQUILA}}},\ \bibinfo {howpublished}
  {\url{https://www.mathworks.com/matlabcentral/fileexchange/3344-2d-schroedinger-poisson-solver-aquila}}
  (\bibinfo {year} {2009--2020})\BibitemShut {NoStop}%
\end{thebibliography}
\end{document}